\documentclass[a4paper,11pt]{article}
\usepackage{jcappub}
\usepackage{array}
\usepackage{multirow}
\usepackage{soul}
\usepackage[dvipsnames]{xcolor}

\graphicspath{{figures/}}

\title{\boldmath $B$-sure I: Minkowski functionals as robustness test for tensor-to-scalar ratio detection from CMB observations}

\author[*]{Claudio Ranucci,\note[*]{Corresponding author.}}
\author{Alessandro Carones,}
\author{L\'eo Vacher,}
\author{Nicoletta Krachmalnicoff}
\author{and Carlo Baccigalupi}
\affiliation{\it International School for Advanced Studies (SISSA),\\Via Bonomea 265, Trieste 34136, Italy}
\affiliation{\it Istituto Nazionale di Fisica Nucleare (INFN) Sezione di Trieste,\\Via Valerio 2, Trieste 34127, Italy}
\affiliation{\it Institute for Fundamental Physics of the Universe (IFPU),\\Via Beirut 2, Trieste 34151, Italy}

\emailAdd{cranucci@sissa.it}

\abstract{The detection of primordial $B$-mode polarisation of the Cosmic Microwave Background (CMB) is a major observational goal in modern Cosmology, offering a potential window into inflationary physics through the measurement of the tensor-to-scalar ratio $r$. However, the presence of Galactic foregrounds poses significant challenges, possibly biasing the $r$ estimate. In this study we explore the viability of using Minkowski functionals (MFs) as a robustness test to validate a potential $r$ detection by identifying non-Gaussian features associated with foregrounds contamination. To do so, we simulate sky maps as observed by a \emph{LiteBIRD}-like CMB experiment, with realistic instrumental and foregrounds modelling. The CMB $B$-mode signal is recovered through blind component separation algorithms, and the obtained (biased) value of $r$ is used to generate Gaussian realisation of CMB signal. Their MFs are then compared with those computed on maps contaminated by foreground residual left by component separation, looking for a detection of non-Gaussianity. Our results demonstrate that, with the experimental configuration considered here, MFs can not be reliably adopted as a robustness test of an eventual $r$ detection, as we find that in the majority of the cases MFs are not able to raise significant warnings about the non-Gaussianity induced by the presence of foreground residuals. In the most realistic and refined scenario we adopted, the test is able to flag non-Gaussianity in $\sim 26\%$ of the simulations, meaning that there is no warning on the biased tensor-to-scalar ratio in $\sim 74\%$ of cases. These results suggest that more advanced statistics than MFs must be considered to look for non-Gaussian signatures of foregrounds, in order to be able to perform reliable null tests in future CMB missions.}


\notoc

\begin{document}

\maketitle
\flushbottom

\section{Introduction}
\label{sec:intro}
Observations of the Cosmic Microwave Background (CMB) have played a crucial role in establishing the $\Lambda$ Cold Dark Matter ($\Lambda$CDM) as the standard cosmological model, providing insights into the origin of the structure, content and evolution of the Universe \cite{WMAP13_results, PlanckVI20_params, ACT_DR4_20}. Most of the information regarding the early phases of the cosmic history has been extracted from temperature fluctuations; however, the fainter CMB polarisation field can still shed new light on the physics of the primordial Universe. A phase of exponential expansion which occurred $10^{-36}-10^{-34} \, \text{s}$ after the Big Bang, called cosmic inflation \cite{Brout78, Starobinsky80, Guth81}, is canonically added to the standard model to explain some features of the hot Big Bang Cosmology. This expansion stretched quantum fluctuations of the metric (primordial gravitational waves, GWs) to cosmological scales, which are expected to produce a specific signature in the CMB polarisation: the primordial $B$-mode pattern \cite{Kamionkowski97, Hu97, Seljak_Zaldarriaga97_gw}. Detecting primordial $B$-modes in CMB polarisation could enable the estimation of the amplitude of primordial gravitational waves, usually quantified by the tensor-to-scalar ratio parameter $r$ \cite{Planck20_inflation}, potentially providing experimental evidence for the inflationary scenario. Due to the faintness of such a signal ($>10^{3}$ times weaker than temperature anisotropies), there is no direct evidence for primordial $B$-modes yet, and current CMB experiments have placed tight constraints on their amplitude, finding $r_{0.05} < 0.032$ \cite{Tristram22} and $r_{0.01} < 0.028$ (with a free-to-vary tensor spectral tilt) \cite{Galloni23} at 95\% confidence when evaluated at a pivot scale of $0.05$ or $0.01 \, \mathrm{Mpc}^{-1}$. At the same time, these experiments also observed that the power spectrum of primordial scalar perturbations, generated by inflation, is not exactly scale-independent, with the scalar spectral index $n_s - 1 \sim - 0.035$ (e.g., \cite{PlanckVI20_params}). This measurement is compatible with several classes of inflationary models predicting $r$ to be in the $10^{-3} - 10^{-2}$ range (see \cite{Kamionkowski_Kovetz16} and references therein).

The detection of primordial $B$-modes thus represents one of the main goals of future CMB missions. Their power spectrum is characterised by the presence of two different features: the reionisation bump ($\ell \lesssim 10$), associated with the scattering of CMB photons with free electrons released during cosmic reionisation, and the recombination bump ($\ell \sim 80$), which corresponds to the imprint of primordial tensor perturbations at the recombination epoch. The first can be measured only through full sky observations from space, while the second can be targeted by observing smaller regions of the sky from the ground \cite{LiteBIRD23PTEP, BICEP21, SO19, S419}.

Even though the only source of primordial large-scale $B$-modes are tensor fluctuations (at linear order), a practical measurement is complicated by several factors. The gravitational deflection of the background CMB photons by the cosmic large-scale structure creates coherent sub-degree distortions in the CMB, known as CMB lensing \cite{Lewis06_lensing}. Through this mechanism, a fraction of the parity-even $E$-modes is transformed into parity-odd $B$-modes at intermediate and small scales \cite{Zaldarriaga_Seljak98_lensing}. Lensing $B$-modes have already been measured by SPTpol \cite{SPT15}, ACTpol \cite{ACT17_lensing}, PolarBear \cite{Polarbear14} and BICEP2/Keck \cite{BICEP18} experiments. \\
Diffuse Galactic emission is significantly polarised, and in particular components such as synchrotron radiation and thermal emission from dust grains produce $B$-modes with a significant amplitude. These two emission mechanisms are obviously prominent around the Galactic plane, but are also clearly detectable at higher latitudes \cite{Skalidis18}. Current measurements of Galactic emission demonstrate that the Galactic $B$-mode signal is dominant over the cosmological signal on all scales \cite{PlanckX16,PlanckXXX16_PIP_dust,PlanckIV20_compsep,PlanckXI20_dust}. At the minimum of polarised Galactic thermal dust and synchrotron, around 80 GHz, their $B$-mode signal represents an effective tensor-to-scalar ratio with amplitude larger than the sensitivity targeted by future CMB experiments, even in the cleanest regions of the sky \cite{Krach16_fgs}.

Component separation methods, which exploit the different spectral energy distributions (SED) of the CMB and foregrounds to separate the different components, are thus of vital importance \cite{Delabrouille07,Leach08}. Practical implementations of these methods must also be able to carry out this separation in the presence of instrumental noise and systematic effects \cite{Abitbol21}. Several component separation methods have been developed in the last years for CMB data analysis; here we mention two categories: 1) parametric-fitting methods \cite{Eriksen08, Stompor08, Vacher22_moments}, which recover the CMB signal by fitting a model of the various sky components; and 2) the ``blind'' methods \cite{Delabrouille09_nilc, Vio08, Carones23_mcnilc}, whose purpose is to recover a cleaned CMB blackbody signal, without any assumption on the SED of foreground emission. Methods of the latter class, in most cases, are also referred to as minimum-variance techniques, since they reconstruct the CMB signal as the minimum variance solution from the linear combination of multi-frequency observations, thus maximally reducing the foreground contamination in the 2-point statistics \cite{Tegmark03}.

The application of component separation algorithms is able to reduce most of the foregrounds impact, but residual contamination in the reconstructed CMB signal may still be comparable in amplitude to the primordial $B$-mode signal to be measured, thus biasing any estimate of the tensor-to-scalar ratio. In recent years, many works have tackled this problem and forecasted the quality of an $r$ estimation with different ground-based and satellite experiments (e.g., \cite{Alonso17_forecasts,LiteBIRD23PTEP,Carones23_nilc}). In general, these works have highlighted how, if left untreated, systematic residuals arising from a simplistic characterisation of foregrounds could bias a $r \sim 10^{-3}$ measurement by several $\sigma$, consequently leading to a false detection, with the outcome of the BICEP2 analysis \cite{BICEP14} being a clear example. Thus, it is crucial to develop methodologies able to detect such residual contamination in the data, in order to validate any potential detection of the tensor-to-scalar ratio.

Minkowski functionals (MFs) are mathematical tools used to assess the statistics of CMB anisotropies; they have been widely explored in the mathematical literature (\cite{Tomita86, Coles87, Gott90}), and have become popular in CMB analysis after the discussion by \cite{Schmalzing98}. In the case of scalar Gaussian isotropic fields such as the CMB temperature anisotropies, the theoretical predictions of these functionals can be accurately computed. Given their low variance, any deviation from Gaussianity or statistical isotropy can be detected at high significance in a model-independent manner. MFs have been adopted to investigate non-Gaussianity and anisotropy in the CMB \cite{PlanckVII20_isotropy}, to study the non-Gaussianity of Galactic signals \cite{Rahman21}, and also to constrain cosmological parameters \cite{Zurcher21}. They are additive for disjoint regions of the sky and invariant under rotations and translations. Moreover, these tools can be applied to blindly detect contamination by foreground and instrumental systematics in the reconstructed CMB map, given their highly non-Gaussian and anisotropic nature. The impact of such systematic and astrophysical effects on the constraining power for primordial non-Gaussianity in temperature data has also been explored by \cite{Ducout13}. Until now, MFs have been largely applied to the scalar fields associated with the CMB polarisation, $E$- and $B$-modes (\cite{Ganesan15, PlanckXVI16_isotropy, Santos16, PlanckVII20_isotropy}), and recently the formalism has been extended to spin-$s$ fields \cite{CarronDuque24_MFs}, including their combination into the polarisation modulus $P^2$ \cite{Carones24_MFs}. The aim of this work is to explore the application of MFs as high-order statistical tools able to test the robustness of a potential detection of the tensor-to-scalar ratio $r$. We investigate if MFs are powerful enough to recognise a contribution to $r$ partially or entirely sourced by residual foreground contamination, through a measurement of non-Gaussianity in the CMB maps.

The paper is organised as follows: in Section~\ref{sec:mfs}, we introduce the MFs theoretical framework, while the full methodology adopted is described in Section~\ref{sec:method}; results are presented in Section~\ref{sec:results}, and we summarise our conclusions and provide future prospects in Section~\ref{sec:conclusions}.

\section{Minkowski functionals}
\label{sec:mfs}
Minkowski functionals (MFs) on the sphere are now well-established tools for CMB data analysis \cite{PlanckVII20_isotropy}; here we recall only the basis of the theory to facilitate comparison with other works. For the complete derivation of the equations and a more detailed discussion, we refer to \cite{CarronDuque24_MFs, Carones24_MFs} and references in therein.

Given a scalar random field $f$ observed on the unit sphere $\mathbb{S}^2$ (e.g., CMB temperature anisotropies, $E$-/$B$-modes), the excursion set at a threshold $u$ is defined as
\begin{equation}
    A_u\left( f, \, \mathbb{S}^2 \right) = \{x \in \mathbb{S}^2 : f(x) > u\}.
    \label{eq:excursion}
\end{equation}
We will omit the arguments of $A_u$ when $f$ and its domain are clear from the context. As recalled in \cite{Schmalzing98}, the morphological properties of these excursion sets can be summarised by the three MFs defined as follows:
\begin{equation}
    \begin{aligned}
        V_0(A_u) &= \int_{A_u} dx, \\
        V_1(A_u) &= \frac{1}{4} \int_{\partial A_u} dr, \\
        V_2(A_u) &= \frac{1}{2\pi}\int_{\partial A_u} \kappa(r) dr,
    \end{aligned}
    \label{eq:Vs}
\end{equation}
where $dr$ denotes a line element along the boundary $\partial A_u$ of the excursion set and $\kappa(r)$ denotes the geodesic curvature of the said boundary. Thus, $V_0$ represents the excursion area, $V_1$ is one-fourth of the boundary length (perimeter), and $V_2$ (genus) is associated with the number of connected regions minus the number of holes.
It is notationally more convenient to replace MFs with the equivalent notion of Lipschitz-Killing curvatures, expressed as
\begin{equation}
    \begin{aligned}
        \mathcal{L}_2 (A_u) &= V_0 (A_u) \\
        \mathcal{L}_1 (A_u) &= 2 V_1 (A_u) \\
        \mathcal{L}_0 (A_u) &= V_2(A_u) + \frac{1}{2\pi} V_0(A_u).
    \end{aligned}
    \label{eq:curvatures}
\end{equation}
For an isotropic Gaussian field (normalised to have unit variance), the expected value of the Lipschitz-Killing  curvatures on the sphere is given by the Gaussian kinematic formula (GKF):
\begin{equation}
    \mathbb{E} \left[ \mathcal{L}_j (A_u) \right] = \sum_{k=0}^{2-j} \frac{\omega_{k+j}}{\omega_k \omega_j} \binom{k+j}{k} \cdot \rho_k (u) \cdot \mathcal{L}_{k+j}(\mathbb{S}^2) \cdot \mu^{k/2},
    \label{eq:gkf}
\end{equation}
where we defined the ``flag'' coefficients \cite{Adler09} as $\omega_j = \pi^{j/2} \left[ \Gamma (\frac{j}{2} + 1)\right]^{-1}$ with $\Gamma(z)$ being the usual Gamma function, and $\mu$ as the derivative of the covariance function at the origin for the field $f$, computed as
\begin{equation}
    \mu = \sum_\ell \frac{\ell(\ell + 1)}{2} \frac{2\ell + 1}{4\pi} C_\ell,
    \label{eq:mu}
\end{equation}
where $C_\ell$ denotes, the angular power spectrum of the field $f$. The quantity $\rho_k$ represents a set of ``density'' functions,
\begin{equation}
    \rho_k (u) = \frac{1}{(2 \pi)^{k/2}} \frac{1}{\sqrt{2 \pi}} \exp \left(- \frac{u^2}{2} \right) H_{k-1}(u) \, ,
    \label{eq:rho}
\end{equation}
with $H_k(u)$ denoting the sequence of Hermite polynomials.
With Eqs.~\eqref{eq:curvatures} and \eqref{eq:gkf} we get the well-known results for the expected values of MFs, which we report normalised by area:
\begin{equation}
    \begin{aligned}
        \frac{\mathbb{E} (V_0)}{4 \pi} & = 1 - \Phi(u) \\
        \frac{\mathbb{E} (V_1)}{4 \pi} & = \frac{1}{8} \exp \left(- \frac{u^2}{2} \right) \mu^{1/2} \\
        \frac{\mathbb{E} (V_2)}{4 \pi} & = \frac{1}{\sqrt{(2 \pi)^3}} \mu \cdot \exp \left(- \frac{u^2}{2} \right) \cdot u.
    \end{aligned}
    \label{eq:mfs}
\end{equation}
where $\Phi$ is the cumulative function of the normal distribution. We note that the first MF depends only on $u$, not on the field itself as long as it is normalised to have unit variance. The above set of equations tell us that the MFs expected values, in the case of a Gaussian random field, depend on the combination of four different independent ingredients:
\begin{enumerate}
    \item a set of universal coefficients (called ``flag'' coefficients);
    \item the Lipschitz-Killing curvatures evaluated on the original manifold (in our case, the unitary sphere);
    \item a power of the derivative of the covariance function of the field at the origin;
    \item a set of ``density'' functions $\rho_k$, dependent only on the threshold at which the MFs are evaluated, and on the statistics of the field $f$.
\end{enumerate}
We note that the presence of a mask only affects the manifold where the map is defined, therefore having an impact only on the MFs normalisation, at leading order.

A key point to notice is that, from Eq.~\eqref{eq:gkf}, the MFs expected value depends both on the level of Gaussianity of the field (in general, on the statistics of the field through $\rho_k$), and on its spectral features through the $C_\ell$ term in the derivative of the covariance function, in Eq.~\eqref{eq:mu}. This implies that two fields will have different MFs values either due to the different statistical properties or to the different shape of the spectra. This point will be discussed in more details in the following Sections.

In this work, MFs of CMB maps are evaluated with \texttt{Pynkowski}\footnote{\href{https://javicarron.github.io/pynkowski/pynkowski.html}{https://javicarron.github.io/pynkowski/pynkowski.html}}, a \texttt{python} package developed to compute MFs and other higher order statistics of fields defined on the sphere and other manifolds. The package has been presented and validated in \cite{CarronDuque24_MFs, Carones24_MFs}, both for scalar ($T,\ E,\ B,\ P^2$) and polarisation ($Q$, $U$) maps.

Following what has been done in previous works \cite{PlanckVII20_isotropy, Carones24_MFs} in order to characterise the MFs we consider two approaches for the scale-dependent analysis of the $B$-mode sky maps: \textit{i)} simultaneous exploitation of all the scales sampled in the maps (i.e., using all the multipoles available from $\ell_\mathrm{min}$ to $\ell_\mathrm{max}$) and \textit{ii)} scales separation by using needlet bands, to isolate the contribution from different ranges of multipoles. The localisation properties of needlets (in real and harmonic space) allow a much more precise, scale-by-scale, interpretation of any possible anomalies, since non-Gaussian or statistically anisotropic signal may be relevant only in a specific range of multipoles (e.g., Galactic residuals on large scales). The needlets approach should then provide a complete view about the nature of possible non-Gaussian features detected with the MFs. Their formalism is described in more detail in Section~\ref{sec:compsep}.

\section{Methodology}
\label{sec:method}

\begin{figure}
    \centering
    \includegraphics[width=1.0\textwidth]{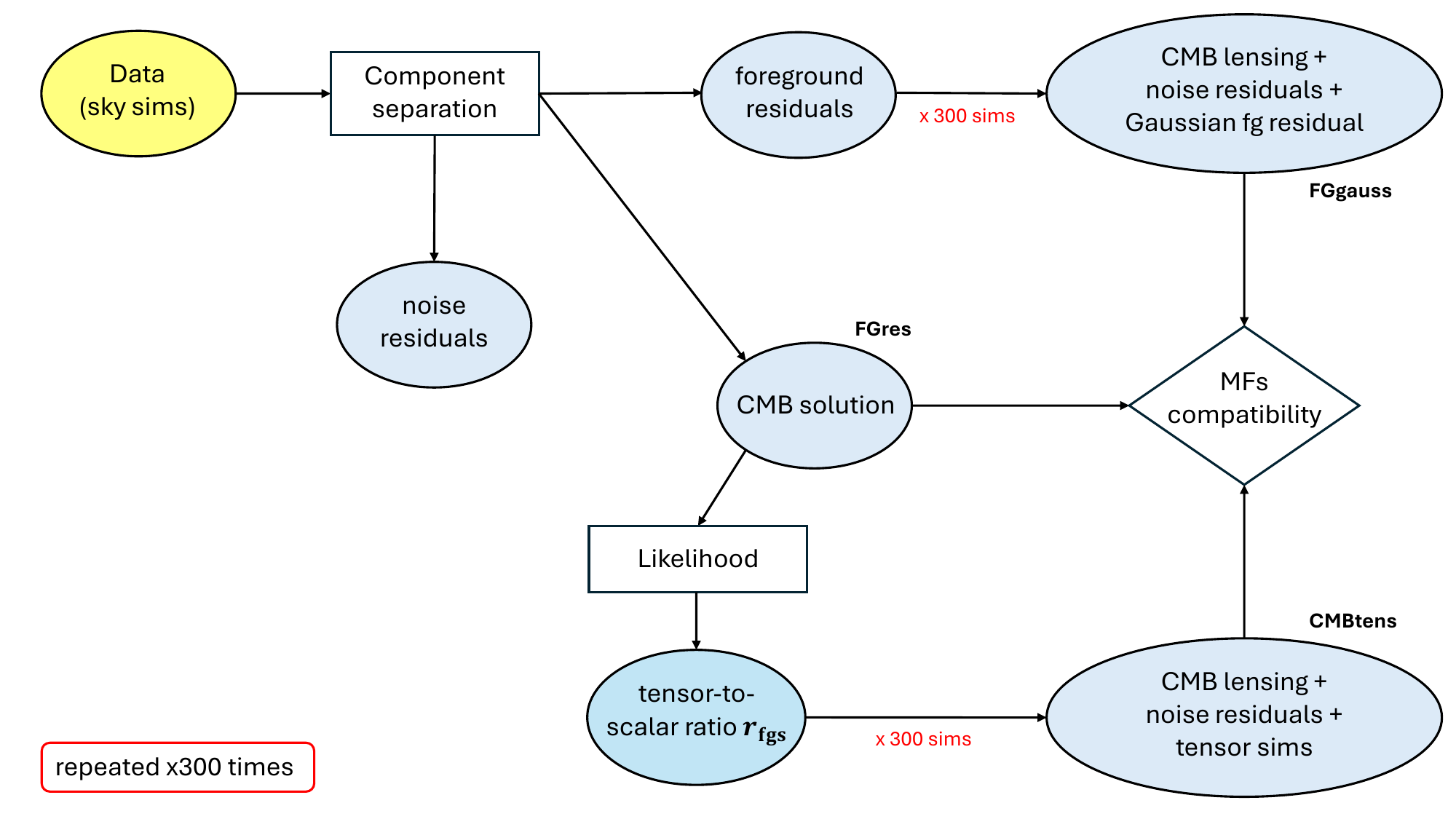}
    \caption{Overview of the steps adopted to build the MFs robustness test. Squares represent pipeline stages, while ovals indicate data products (input in yellow, intermediate in light blue). This procedure is then repeated for 300 ``data" simulations.}
    \label{fig:flowchart}
\end{figure}

As outlined in Section~\ref{sec:intro}, the goal of this work is to build a robustness test for a potential tensor-to-scalar ratio detection, using MFs as statistical tools.
For reference, we describe here all the steps we follow to build this MFs-based robustness test, with the proposed analysis applied to the instrumental configuration of a \emph{LiteBIRD}-like experiment\footnote{\href{https://www.isas.jaxa.jp/en/missions/spacecraft/future/litebird.html}{\emph{LiteBIRD}}: Lite (Light) satellite for the study of $B$-mode polarization and Inflation from cosmic background Radiation Detection.}. The procedure is also illustrated in Figure~\ref{fig:flowchart}. Each one of the item is then discussed in details in the correspondent Section:
\begin{enumerate}
    \item first, we generate a set of multi-frequency \textit{LiteBIRD}-like full-sky simulations in polarisation, ($Q$ and $U$ Stokes parameters), by co-adding CMB (with $B$-modes sourced by Gaussian realisation of lensing only, $r = 0$), instrumental noise, and different models of polarised emission from Galactic foregrounds (Section~\ref{sec:sims}). Simulations are then converted to $B$-mode maps through a full-sky harmonic transformation;
    \item we apply component separation to these $B$-mode maps, obtaining three outputs: a cleaned CMB map (CMB solution), a map of noise residuals and a map of (non-Gaussian) foreground residuals (Section~\ref{sec:compsep}), with the CMB solution containing the other two maps;
    \item we mask in each product the expected most foreground-contaminated sky regions. We evaluate the likelihood function on the power spectrum of the cleaned CMB map, inferring the tensor-to-scalar ratio $r_\text{fgs}$, eventually biased by foregrounds (Section~\ref{sec:likelihood});
    \item we use the measured $r_\text{fgs}$ to generate 300 Gaussian realisations of CMB $B$-mode polarisation maps sourced by primordial GWs only, $C_\ell^{BB} = r_\text{fgs} \cdot C_\ell^{r = 1}$ (Section~\ref{sec:sims_tens});
    \item using the power spectrum of the foreground residuals map derived from component separation (step 2), we generate a set of 300 realisations of Gaussian B-mode maps from it (Section~\ref{sec:sims_gauss_fgs});
    \item we compute MFs of the following sets of maps:
    \begin{itemize}
        \item \texttt{FGres}: co-addition of CMB with $r = 0$, noise residuals, non-Gaussian foreground residuals (i.e., the CMB solution returned by the component separation, step 2);
        \item \texttt{CMBtens}: co-addition of CMB with $r = r_\text{fgs}$ (step 4) and noise residuals after component separation;
        \item \texttt{FGgauss}: co-addition of CMB with $r = 0$, noise residuals after component separation and Gaussian foreground residuals (step 5);
    \end{itemize}
    all datasets include the input realisations of the lensing signal in the CMB component;
    \item we run a statistical analysis to assess if MFs are able to robustly detect deviations between the cleaned CMB map and the Gaussian simulations;
\end{enumerate}
This procedure is repeated for 300 realisations of the input multi-frequency maps, and applied to different combination of component separation pipeline and foregrounds model.

\subsection{Input simulations}
\label{sec:sims}
In this Section we describe the sky simulations that we use as input for our pipeline (step 1), while the CMB tensor and Gaussian foregrounds simulations outlined in steps 4 and 5 are described in Section~\ref{sec:sims_tens} and \ref{sec:sims_gauss_fgs}, respectively. Maps generation, manipulation, and analysis is carried out adopting the \texttt{HEALPix}\footnote{\href{https://healpix.jpl.nasa.gov/}{https://healpix.jpl.nasa.gov/}} \cite{healpix} pixelisation scheme by means of the \texttt{healpy}\footnote{\href{https://healpy.readthedocs.io/en/latest/}{https://healpy.readthedocs.io/en/latest/}} package \cite{healpy}. For all our simulations, the pixel resolution is defined by $N_\text{side} = 64$, corresponding to an angular resolution of about 55 arcmin, with the maximum multipole considered being $\ell_\text{max} = 3 N_\text{side} - 1 = 191$.

CMB maps are generated as Gaussian realisations from the angular power spectra of the best-fitting \emph{Planck} 2018 parameters \cite{PlanckVI20_params}; we produce $N = 300$ simulations where $B$-modes are sourced by gravitational lensing only, with no primordial signal ($r = 0$). The adopted \emph{LiteBIRD} specifications are taken from the latest forecast of the experiment (Table 13 in \cite{LiteBIRD23PTEP}), planned to cover 15 frequency bands with 22 effective channels between 40 and 402 GHz. All CMB maps are smoothed to a common angular resolution of $\text{FWHM}_\text{out} = 70.5 \, \mathrm{arcmin}$, which is the largest beam considered (corresponding to the 40 GHz channel).

Noise maps are simulated by generating Gaussian realisations of white noise at the different frequency channels; noise maps are smoothed to a common resolution by considering an effective beam as
\begin{equation}
    \text{FWHM}_\text{eff}(\nu) = \sqrt{\text{FWHM}_\text{out}^2 - \text{FWHM}(\nu)^2}.
    \label{eq:fwhm_eff}
\end{equation}

To simulate Galactic foregrounds, we use the Python Sky Model (\texttt{PySM})\footnote{\href{https://pysm3.readthedocs.io/en/latest/index.html}{https://pysm3.readthedocs.io/en/latest/index.html}} \cite{PySM,Zonca21_PySM,Panexp25_PySM}, a \texttt{python} package that generates full-sky simulations of Galactic foregrounds in intensity and polarisation relevant for CMB experiments. Since we are focusing on polarisation, we are only interested in thermal dust (\texttt{d}) and synchrotron (\texttt{s}) emission, as they are the main contaminants of observations.

In this work we adopt three different models for synchrotron and thermal dust emission, of increasing complexity: \texttt{d0s0}, \texttt{d1s1}, and \texttt{d10s5}. The former one is a simplistic view of the sky that we use for testing and validation of our pipeline, while the others are models closer to realistic observations. We refer to the \texttt{PySM} papers and documentation for more details.

We generate $Q$/$U$ full-sky simulations for each of the three models, one per frequency channels; each map is smoothed to 70.5 arcmin and then co-added to the CMB and noise simulations. We end up with sky simulations which include CMB, noise, and foregrounds, with 300 different realisations of CMB and noise for each of the three considered \texttt{PySM} models.

\subsection{Component separation}
\label{sec:compsep}
As mentioned in Section~\ref{sec:intro}, in this work we adopt minimum-variance algorithms as our component separation methods to recover CMB $B$-modes from multi-frequency simulations of the sky. In particular, we use two implementations: Needlet Internal Linear Combination (NILC) and Multi-Clustering NILC (MC-NILC). NILC has already been applied in the past for CMB data analysis (e.g., \emph{WMAP} \cite{Basak11_nilc_temp}, \emph{Planck} \cite{PlanckIV20_compsep}) and both NILC and MC-NILC will be used for current and future CMB surveys (\emph{SO} \cite{SO19,Wolz24}, \emph{LiteBIRD} \cite{LiteBIRD23PTEP}).

NILC belongs to the category of blind component separation methods, since it aims at reconstructing the CMB signal without any assumptions on the foreground spectral properties. It thus represents a valuable alternative to parametric approaches, as it is not affected by spectral mis-modelling of the Galactic polarised emission, which may significantly bias the final estimate of the tensor-to-scalar ratio.

In a framework with complex foreground emission, the NILC pipeline may present some limitations or be suboptimal, since it performs simple local variance minimisation across the sky, without fully handling the local spectral variations of foregrounds. MC-NILC \cite{Carones23_mcnilc} is a NILC extension which performs component separation independently in different sky regions, where the polarised $B$-mode Galactic emission shows similar spectral properties. The spatial variability of dust and synchrotron spectral parameters is assessed by identifying a blind tracer of their distribution across the sky. This tracer is obtained from the simulated dataset by applying a model-independent and realistic de-noising technique \cite{Carones23_mcnilc}.

\begin{figure}
    \centering
    \includegraphics[width=0.45\textwidth]{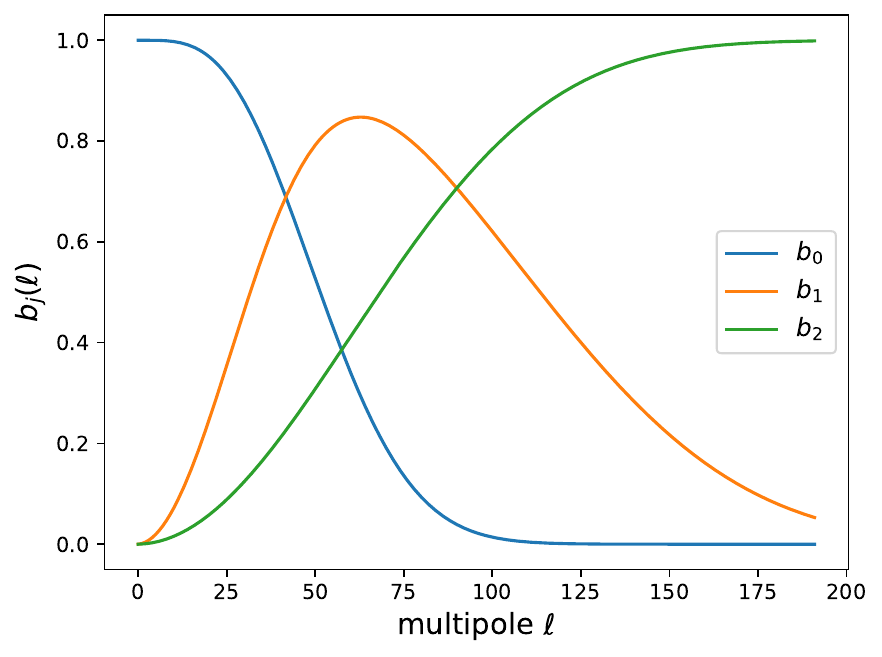}
    \includegraphics[width=0.45\textwidth]{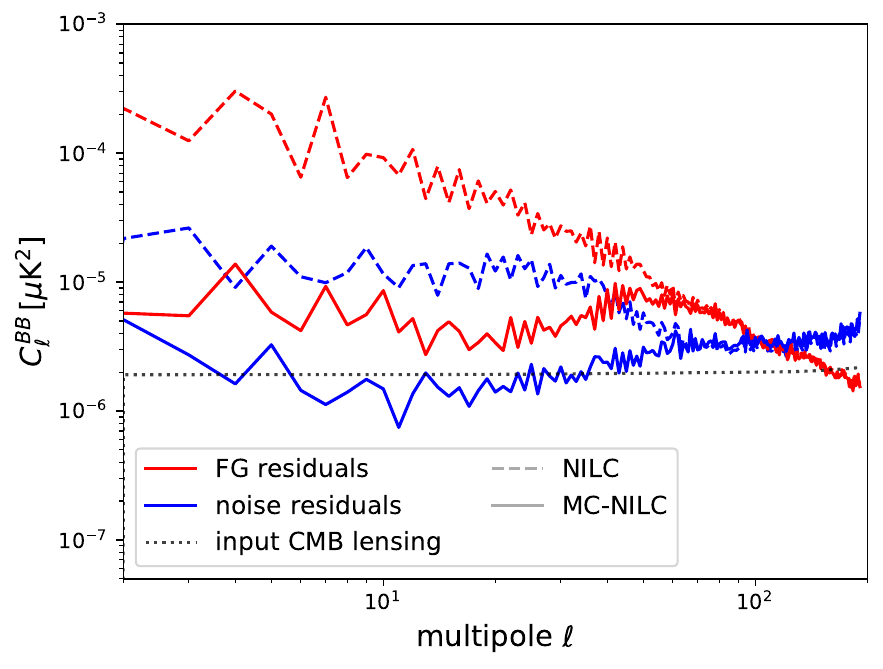}
    \caption{\textit{Left panel}: needlet bands configuration adopted in this work, where the first bands have been merged. \textit{Right panel}: full-sky angular power spectra evaluated on foreground (red) and noise (blue) residuals, for NILC (dashed lines) and MC-NILC (solid lines), applied to the \texttt{d10s5} model. Black dotted line is the input CMB spectrum (lensing-only).}
    \label{fig:bands_spectra_nilc_mcnilc}
\end{figure}

\begin{figure}
    \centering
    \includegraphics[width=0.9\textwidth]{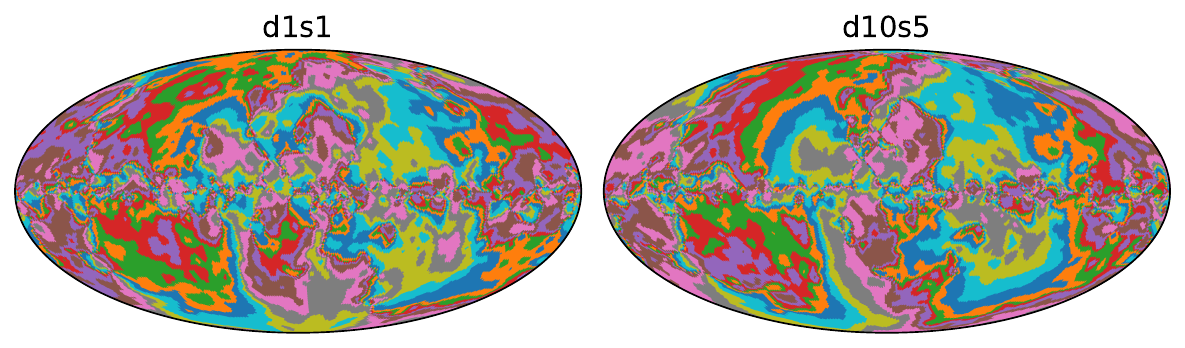}
    \caption{Sky partitions adopted within MC-NILC component separation for \texttt{d1s1} (\emph{left}) and \texttt{d10s5} (\emph{right}) foreground models.}
    \label{fig:tracers}
\end{figure}

\begin{figure}
    \centering
    \includegraphics[width=1.\textwidth]{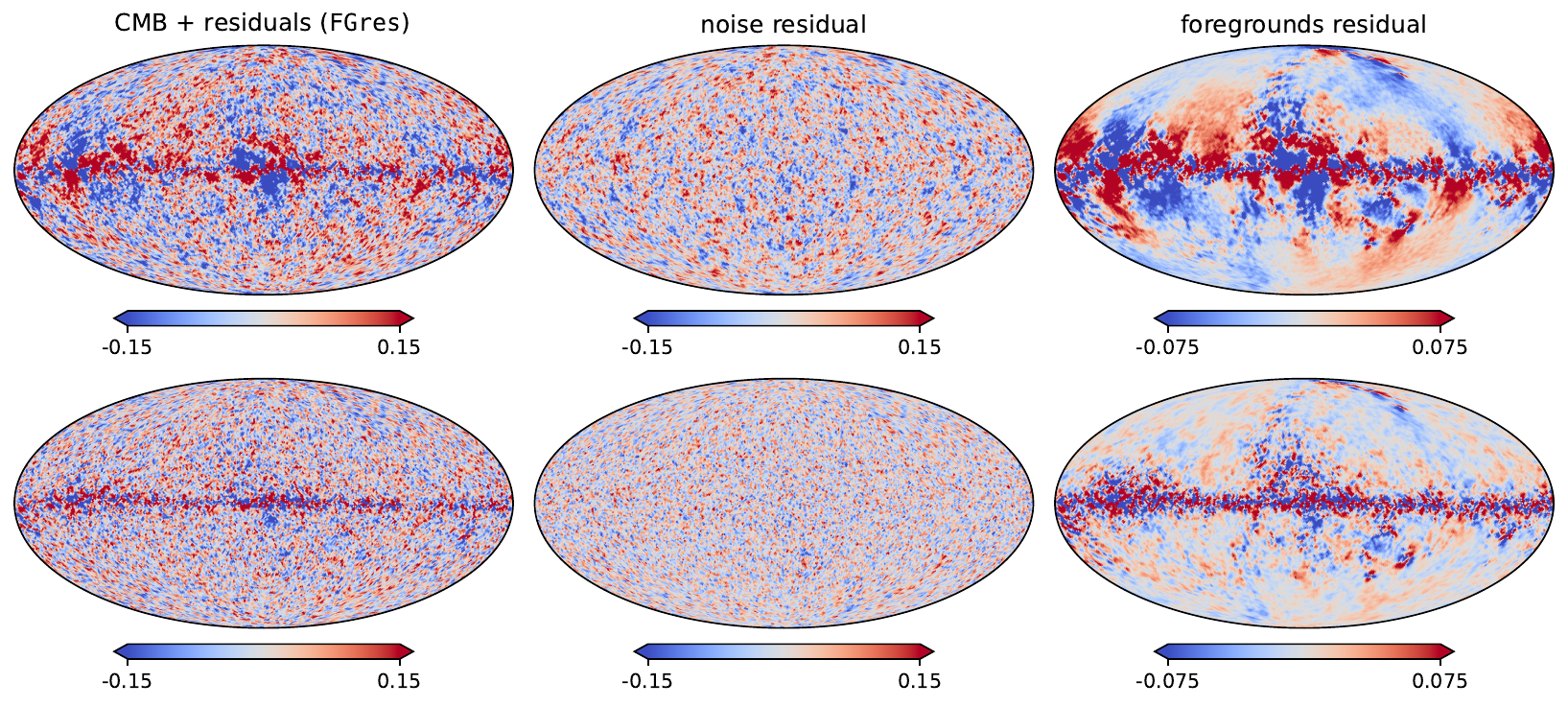}
    \caption{Component separation outputs, for one simulation. \emph{Left column:} CMB solution. \emph{Middle column:} noise residuals. \emph{Right column:} foreground residuals. Outputs are shown for NILC (upper row) and MC-NILC (lower row) methods, applied on the \texttt{d10s5} model. Units are $\mu \mathrm{K_{CMB}}$.}
    \label{fig:compsep_maps}
\end{figure}

Both algorithms relies on the notion of needlets. Needlets are a wavelet system that guarantees simultaneous localisation in harmonic and pixel space; they have been introduced in the statistical literature in \cite{Narcowich06} and first applied to CMB data in \cite{Pietrobon06}. Given any field of spin $s$ defined on the sphere, the corresponding needlet maps $\beta^s_j$ are obtained by filtering its harmonic coefficients $a^s_{\ell m}$ with a harmonic weighting function $b$, which isolates modes at different angular scales for each needlet scale $j$ \cite{Geller10}:
\begin{equation}
    \beta^s_j(\hat{n}) = \sum_{\ell, m} \left[a^s_{\ell m} \cdot b \left( \frac{\ell}{B^j} \right) \right] \,_sY_{\ell m}(\hat{n})
    \label{eq:needlets}
\end{equation}
where $\hat{n}$ is a direction in the sky and $_sY_{\ell m}$ are the spin-weighted spherical harmonics. This procedure in harmonic space is equivalent to performing a map convolution in real domain. The shape of the needlet bands is defined by the choice of the harmonic function $b$, whose width is set by the parameter $B$: lower values of $B$ correspond to a tighter localisation in harmonic space (less multipoles entering into any needlet coefficient). We adopt the Mexican needlet construction \cite{Narcowich06} as our harmonic function, with $B = 1.3$ (Figure~\ref{fig:bands_spectra_nilc_mcnilc}, left panel). The needlet filters have been obtained through the \texttt{MTNeedlet}\footnote{\href{https://javicarron.github.io/mtneedlet/index.html}{https://javicarron.github.io/mtneedlet/index.html}} \cite{CarronDuque19_ps} package.

The adopted MC-NILC sky partitions are illustrated in Figure~\ref{fig:tracers}, while an example of the performances of the two algorithms is reported in Figures~\ref{fig:bands_spectra_nilc_mcnilc} (right panel) and \ref{fig:compsep_maps}. The amount of leftover contamination can be visually compared both at the map and at the power spectrum level, and the improvement achieved by MC-NILC can be clearly seen in the reduction of the noise and foreground residuals. \\
Usually a component separation algorithm returns a single CMB ``solution'' as final product, containing the cleaned CMB signal and some contamination from noise and foregrounds. In a realistic experiment, these residuals maps are not accessible. Since in this analysis we are working with simulated datasets, we have complete control over the separated sky components, as we independently generated CMB, noise, and foregrounds realisations. Component separation weights can be applied to the input noise and foregrounds maps to obtain residuals maps, useful for pipeline validation and for the characterisation of the MFs robustness test.

\subsection{Likelihood}
\label{sec:likelihood}
\begin{figure}
    \centering
    \includegraphics[width=0.95\textwidth]{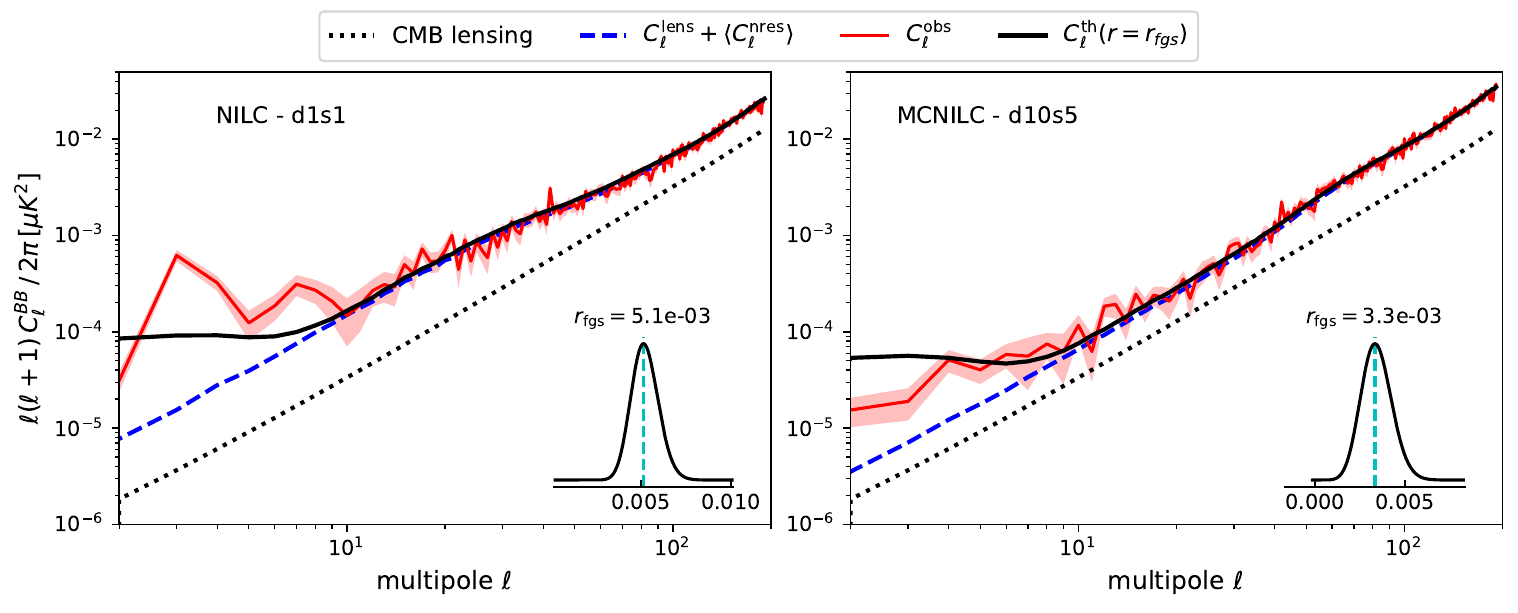}
    \caption{Output of the likelihood evaluation for a single simulation, for the NILC - \texttt{d1s1} (left) and MC-NILC - \texttt{d10s5} (right) scenarios, considering all the multipoles. \emph{Main plot}: power spectra, with $C_\ell^\text{th} (r = 0)$ in dashed blue, $C_\ell^\text{obs}$ in red solid, $C_\ell^\text{lensing}$ in dotted black, and best fit spectrum in black solid. The shaded area represents the standard deviation of $C_\ell^\text{obs}$ across 300 realisations. \emph{Inset}: normalised likelihood on the tensor-to-scalar ratio, with the peak value indicated by the dashed line.}
    \label{fig:spectra_likelihood_pixel}
\end{figure}
In order to infer the tensor-to-scalar ratio from the component separated maps, we adopt an inverse-Wishart distribution as our likelihood \cite{Hamimeche_Lewis09,Gerbino20_likelihood}:
\begin{equation}
    - \ln \mathcal{L} \left( C_\ell^\text{obs} | r \right) = \frac{1}{2} \sum_{\ell} (2\ell + 1) f_\text{sky} \left[ \frac{C_\ell^\text{obs}}{C_\ell^\text{th}(r)} + \ln C_\ell^\text{th}(r) - \frac{2\ell - 1}{2\ell + 1} \ln C_\ell^\text{obs} \right].
    \label{eq:likelihood}
\end{equation}
where $C_\ell^\text{obs}$ is the ``observed'' power spectrum evaluated on the CMB solution map, containing lensing, noise and foreground residuals, and $C_\ell^\text{th}$ is the theoretical $BB$ spectrum built as
\begin{equation}
    C_\ell^\text{th}(r) = C_\ell^\text{lensing} + \langle C_\ell^\text{nres}\rangle + r \cdot C_\ell^{r=1}.
    \label{eq:cl_th}
\end{equation}
In the above equation, $C_\ell^\text{lensing}$ is the $B$-mode spectrum induced by gravitational lensing, $C_\ell^{r=1}$ is the theoretical $B$-mode power spectrum sourced by tensor perturbations only with $r = 1$, and $\langle C_\ell^\text{nres}\rangle$ is the average of the spectra computed on all the other $N - 1 = 299$ noise residuals maps obtained from the component separation.

Angular power spectra are computed with the \texttt{anafast} routine as implemented in the \texttt{healpy} package \cite{healpix,healpy}. To avoid strong contamination from the Galactic plane, a simple mask\footnote{Available at the \href{https://pla.esac.esa.int/\#maps}{Planck Legacy Archive}.} is applied to the maps, retaining a sky fraction of $f_\text{sky} \simeq 60\%$. As it is well-known, estimating power spectra with \texttt{anafast} does not account for correlations among multipoles induced by the presence of a mask (usually referred as mode-coupling \cite{Lewis01_masked}). However, as discussed in \cite{LiteBIRD23PTEP}, this approximation has a negligible impact on the power spectra estimation of foreground and noise residuals over large sky fractions, as considered in this work.

We evaluate the likelihood function over an interval of $r$ values in the range $r \in [-1.5 \times 10^{-4}, 0.03]$, with a constant spacing of $\Delta r = 2 \times 10^{-6}$. Values of $r < 0$ are included in the analysis in order to be sensitive to negative biases on the tensor-to-scalar ratio, as our generated CMB simulations assumes $r = 0$. We take the peak value of the likelihood as our best fit tensor-to-scalar ratio $r_\text{fgs}$, where the $\textbf{fgs}$ label indicates that the bias on the parameter is entirely due to the presence of foreground residuals, which contribute to the power observed on the large scales and mimic the behaviour of a primordial GWs signal.

\begin{figure}
    \centering
    \includegraphics[width=0.95\textwidth]{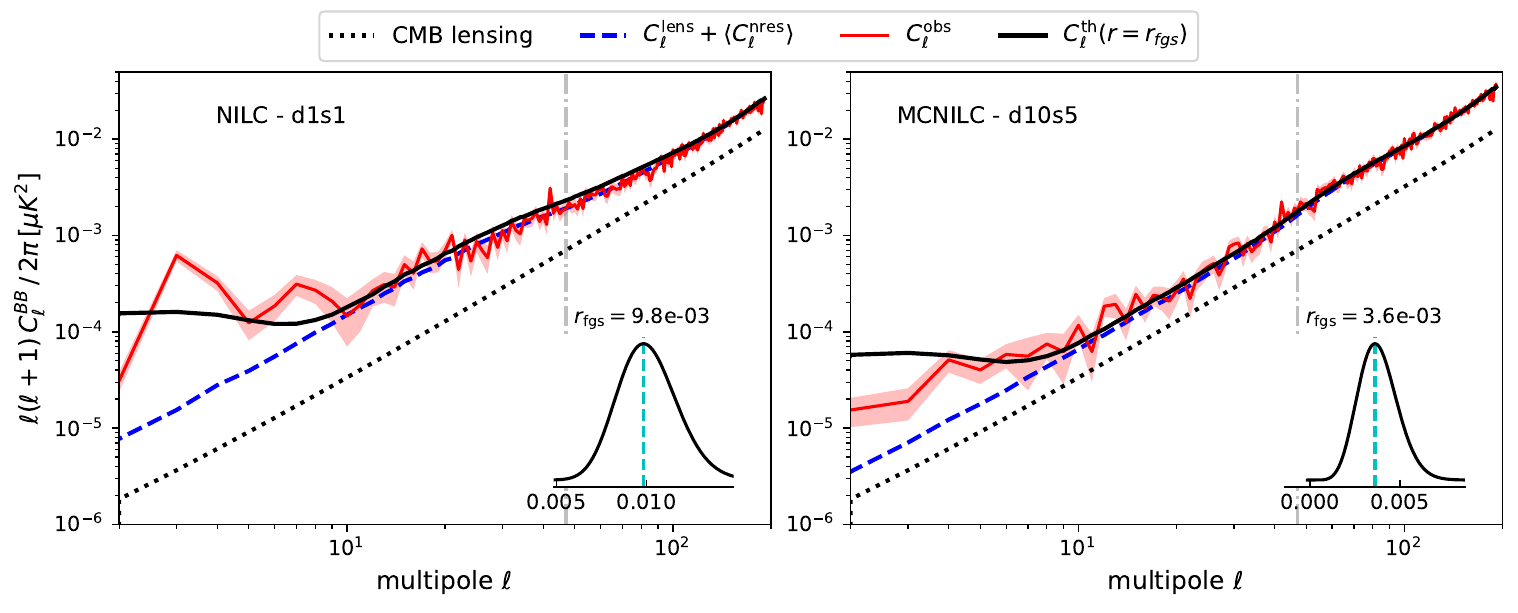}
    \caption{Same as Figure~\ref{fig:spectra_likelihood_pixel}, but considering only the largest scales in the likelihood evaluation. The maximum multipole included is indicated by the vertical line.}
    \label{fig:spectra_likelihood_needlets}
\end{figure}

The inferred value of the tensor-to-scalar ratio gathers contributions from all the angular scales included in the analysis. In general, the largest scales (low $\ell$) contain most of the information about foregrounds and inflation, while the smaller scales (high $\ell$) are dominated by noise and CMB lensing, having therefore a lower impact on the $r$ measurement. To highlight the different contributions, we also evaluate the likelihood function over three multipoles range: $\Delta \ell_0 \in [2, 60]$, $\Delta \ell_1 \in [15, 175]$, $\Delta \ell_2 \in [40, 191]$. These intervals follow the scales sampled by the needlet bands used for component separation and MFs computation, including all multipoles where each of the adopted needlet bands is greater than 30\%. We will focus on the first interval, as we are mainly interested on the impact of foreground residuals on the largest angular scales. An example of the likelihood inputs and outputs is provided in Figure~\ref{fig:spectra_likelihood_pixel}, where we report the output and best-fit tensor $B$-mode angular power spectra together with the corresponding posterior of $r_\text{fgs}$. Figure~\ref{fig:spectra_likelihood_needlets} instead outlines, for the same cases, the best-fit tensor spectra obtained when only the largest scales ($\Delta \ell_0$) are considered in the likelihood analysis.

Following the procedure adopted in a realistic data analysis, we evaluate the goodness-of-fit achieved to understand if we are (already) able to assess whether the bias on $r$ is induced by the presence of foreground residuals. We note that Eq.~\eqref{eq:likelihood} is not a Gaussian likelihood, and thus it is not possible to define a $\chi^2$ statistic in the usual way. However, we can still evaluate the fit quality by relying on simulations, as follows (for each simulation):
\begin{enumerate}
    \item we use the inferred tensor-to-scalar ratio $r_\text{fgs}$ to generate 300 Gaussian random realisations of CMB $B$-mode maps ($C_\ell^{BB} = r_\text{fgs} \cdot C_\ell^{r = 1}$) and we co-add them with simulations of CMB lensing and noise residuals (provided by the component separation). Doing so means taking the CMB solution and replacing the foreground residuals contribution with realisations of tensor perturbations;
    \item for each of these combined CMB lensing, noise, and tensor perturbation maps we compute their power spectrum and we fit for $r$ with the same likelihood function of Eq.~\eqref{eq:likelihood}. We take each minimum of the $- \ln \mathcal{L}$ to build a distribution which we can use to compare with the value obtained by the corresponding best fit $r_\text{fgs}$;
    \item we compute the probability-to-exceed (PTE), that is the amount of the distribution which falls above the value correspondent to the best fit $r_\text{fgs}$. If the PTE is less than 0.05 (95\%), we deem the fit as not good for the considered simulation.
\end{enumerate}
We repeat this analysis for all the 300 input simulations. In general, we expect the PTE distribution to be uniform in order to have a reasonable goodness-of-fit, with the percentage of incompatible simulations that should be $\sim 5\%$. The results of this analysis are shown in Table~\ref{tab:eta_like}, while the PTE distribution for the 300 simulations can be found in Appendix~\ref{sec:appendix}.

\begin{table}
    \centering
    \caption{Percentage of incompatible simulations from the goodness-of-fit analysis, considering all scales at the same time or only the largest scales. Values are reported for each scenario considered.}
    \begin{tabular}{c m{1.5cm} m{1.5cm} m{1.5cm}}
        \multicolumn{4}{c}{all multipoles ($\Delta \ell$=2--191)} \\
        \hline
        & \texttt{d0s0} & \texttt{d1s1} & \texttt{d10s5} \\ \hline
        NILC & 3.7 & 97.7 & 74.0 \\ \hline
        MC-NILC &  & 28.0 & 21.7 \\
        \hline
        \\
        \multicolumn{4}{c}{large-scales only ($\Delta \ell_0$=2--60)} \\
        \hline
        & \texttt{d0s0} & \texttt{d1s1} & \texttt{d10s5} \\ \hline
        NILC & 4.0 & 98.0 & 71.0 \\ \hline
        MC-NILC &  & 51.7 & 42.7 \\
        \hline
    \end{tabular}
    \label{tab:eta_like}
\end{table}

As it can be seen from the Table, in almost all the cases we do not have a good fit quality, with PTE distributions far from being uniform. This is suggesting that the shape of the power spectra at the largest scale do not match enough for the fit to be acceptable: in a realistic scenario, we would likely get a $\text{PTE} < 0.05$, already ruling out the cosmological origin of the inferred tensor-to-scalar ratio. In this analysis we are restricting ourselves to specific foreground models, component separation pipelines, and instrumental configurations, which may not represent the most general case. To illustrate a broader scenario, we also consider a situation where no spectral mismatch is present and only higher-order statistics can reveal discrepancies between residuals and primordial cosmological signal.  We will further describe this point in Section~\ref{sec:sims_gauss_fgs}.

\subsection{Gaussian simulations for MFs comparison}
\label{sec:sims_gauss}
\begin{figure}
    \centering
    \includegraphics[width=1.0\textwidth]{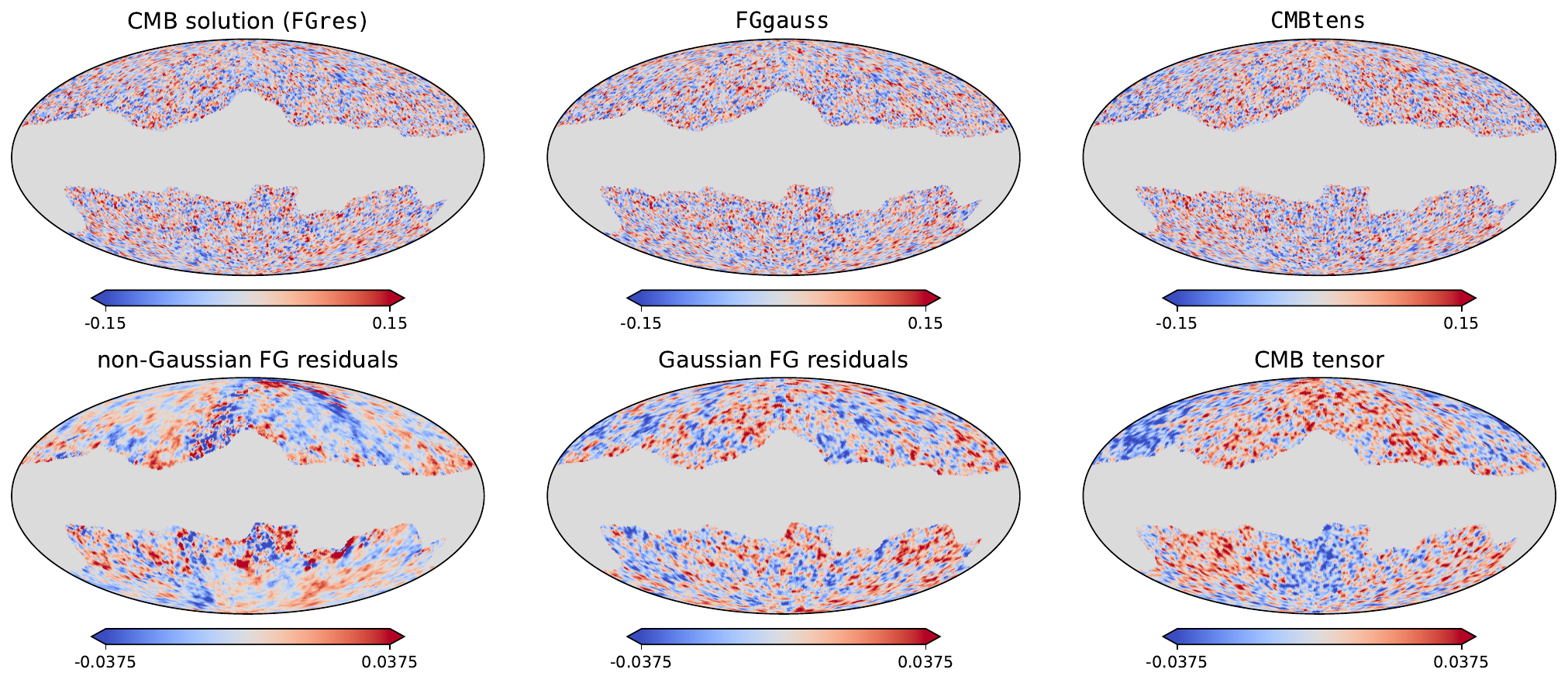}
    \caption{Example of maps of which MFs are computed, for the MC-NILC - \texttt{d10s5} case, considering all the scales. \emph{Upper row:} CMB solution (\texttt{FGres}) on the left, Gaussian simulations (\texttt{FGgauss}, \texttt{CMBtens}) in the middle / on the right. \emph{Lower row:} isolated contributions, without CMB lensing and noise, with non-Gaussian (left) and Gaussian (middle) foreground residuals, and tensor perturbations (right).}
    \label{fig:mfs_maps_pixel}
\end{figure}

\begin{figure}
    \centering
    \includegraphics[width=1.0\textwidth]{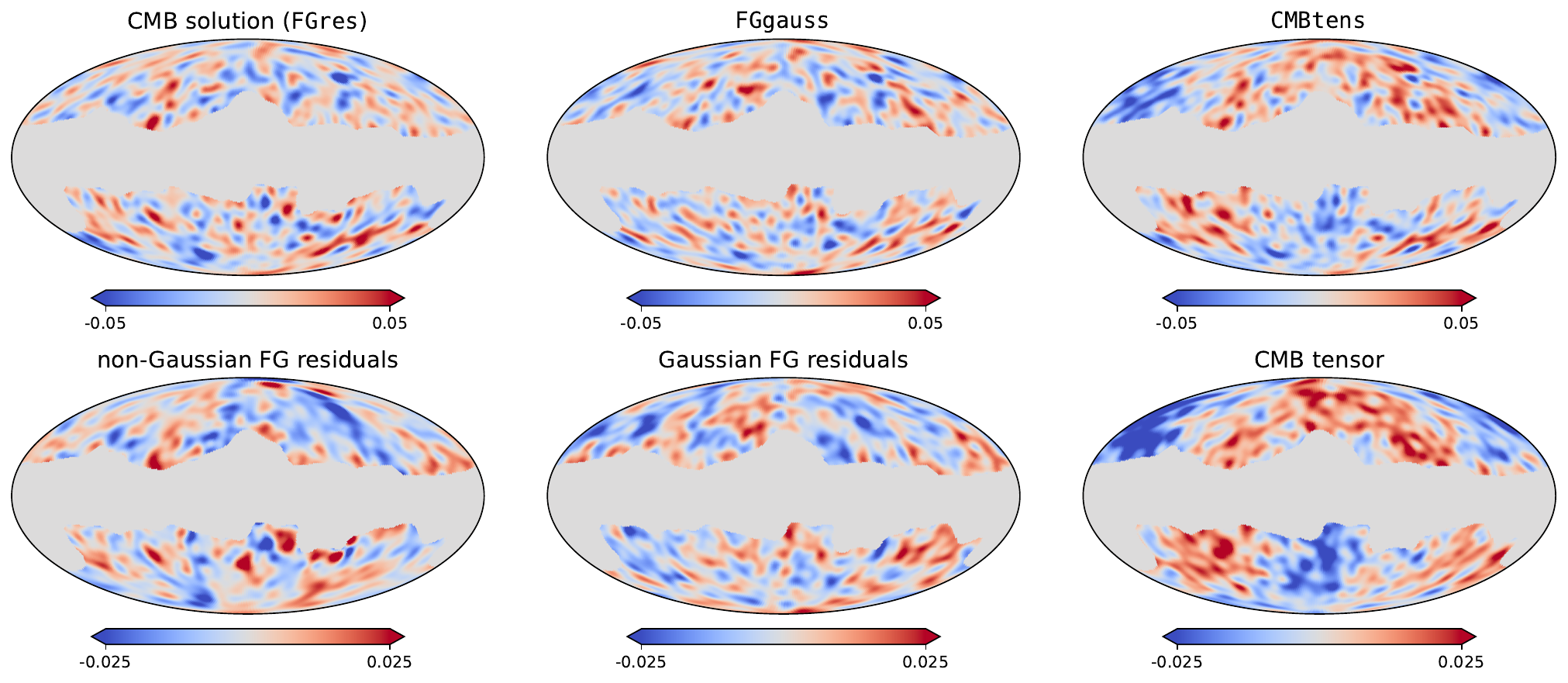}
    \caption{Same as Figure~\ref{fig:mfs_maps_pixel}, but in needlet space (i.e., considering separate scales), for the MC-NILC - \texttt{d10s5} scenario in the $b_0$ needlet band (multipoles $\ell \lesssim$ 60).}
    \label{fig:mfs_maps_needlet}
\end{figure}

The aim of the whole analysis is to compare MFs evaluated on the CMB solution (\texttt{FGres}) against those computed on two different sets of Gaussian simulation: tensor perturbations (\texttt{CMBtens}) and foreground residuals (\texttt{FGgauss}). These Gaussian simulations are co-added with CMB lensing and noise residuals maps, in order to make a comparison with the MFs of \texttt{FGres} maps. Therefore, the compared maps only differ for the foreground residuals part, which is replaced by either tensor perturbations or Gaussian foreground residuals. In this Section we describe how these two sets of Gaussian simulations are generated. An example for one simulation of these two types of maps and their co-addition with CMB lensing and noise residuals is reported in Figures~\ref{fig:mfs_maps_pixel} and \ref{fig:mfs_maps_needlet}.

\subsubsection{Tensor perturbations (\texttt{CMBtens})}
\label{sec:sims_tens}
After the likelihood evaluation on the component-separated maps, we obtain $N = 300$ values (one per simulation) of the best-fit tensor-to-scalar ratio $r_\text{fgs}$, biased by foreground residuals. Since our ultimate goal is to assess the MFs capability to discriminate between foreground contamination and a cosmological signal, we generate a set of Gaussian realisations of tensor perturbations and co-add them with simulated CMB lensing and noise residuals maps. Specifically, for each inferred $r_\text{fgs}$, we generate other 300 realisations of $B$-mode maps (i.e., 300 $\times$ 300 simulations in total) sourced only by primordial GWs ($C_\ell^{BB} = r_\text{fgs} \cdot C_\ell^{r = 1}$), smoothed to the same beam resolution of the component separation outputs. If the detected tensor-to-scalar ratio is of astrophysical origin, then the CMB solution (that includes anisotropic contamination) and the tensor simulations (Gaussian) should be statistically incompatible when observed through a higher-order statistics like MFs, failing the robustness test. Vice versa, if no deviation is observed between two maps, it means that MFs are not able to detect the level of non-Gaussian contribution coming from the foreground residuals. In the latter case, the test is passed and no flag is raised. The complete statistical compatibility analysis is outlined in Section~\ref{sec:compatible}.

\subsubsection{Foreground residuals (\texttt{FGgauss})}
\label{sec:sims_gauss_fgs}
The noise and foreground residuals maps provided by component separation represent the level of contamination left in a cleaned CMB map by the adopted algorithm. The right column of Figure~\ref{fig:compsep_maps} shows how foreground residuals display features of Galactic origin, with evident anisotropic structures over the sky. These structures are not so obvious when foreground residuals are mixed with CMB and noise (left column of Figure~\ref{fig:compsep_maps}). The contamination is stronger along the Galactic plane (always masked in data analysis), while at higher latitudes the foregrounds anisotropy is not strikingly evident.

In Section~\ref{sec:likelihood} we fit the power spectrum of the CMB solution, measuring a biased tensor-to-scalar ratio. We find that the shapes of the power spectra --- foregrounds against tensor perturbations --- do not match, with a bad fit quality (Table~\ref{tab:eta_like}). This introduces a problem in the interpretation of MFs results: as outlined in Section~\ref{sec:mfs}, their expected values depend both on the map statistics and on the spectral features through the $C_\ell$ term in Eq.~\eqref{eq:gkf}. Two maps with same statistics but different spectra will have different MFs values.

We are mainly interested in using MFs to detect deviations from the Gaussian statistics, since a difference in the power spectrum shape would already be detected by a poor goodness-of-fit in the likelihood analysis. To disentangle the two effects, we also evaluate MFs on the \texttt{FGgauss} maps, built by replacing the non-Gaussian foreground residuals of the \texttt{FGres} maps with 300 Gaussian realisation of it. These simulations are produced by taking the power spectrum of foreground residuals outside the adopted mask, and using it to generate Gaussian realisations with the \texttt{synfast} routine from \texttt{healpy}. This means that we are treating the spectrum of a single realisation as the correct underlying theoretical spectrum, with the average spectrum of the Gaussian simulations converging to the non-Gaussian one. In this way, in the comparison between \texttt{FGres} and \texttt{FGgauss}, MFs are evaluated on sky maps with the same ``observed" power spectrum but different statistics, providing a critical benchmark to evaluate the MFs ability in discriminating the two fields. In a realistic experiment, this would translate to an optimal fit in the likelihood step, reproducing a situation where we obtain a biased estimate of the tensor-to-scalar ratio with a good fit quality. This approach also helps in alleviating limitations induced by some assumptions we made in our analysis, such as: \textit{i)} the adoption of a simplified sky mask, which is sub-optimal for mitigating foregrounds contamination; \textit{ii)} the assumption of an inverse-Wishart form for the likelihood, which does not strictly hold on a masked sky, and \textit{iii)} the poor goodness-of-fit in the inference of the tensor-to-scalar ratio. The first two assumptions may contribute to the bias on $r$, and a mask or likelihood optimisation would therefore reduce the bias on the parameter.

We stress that in a realistic situation we do not have access to the component separation residuals and, consequently, to the possibility of generating this kind of simulations. Nonetheless, it is useful to explore the MFs detection power in an idealised scenario where the contribution from non-Gaussianity is isolated, in order to understand if this higher-order statistic can be a reliable tool in validating a tensor-to-scalar ratio potential detection.

\subsection{Compatibility analysis}
\label{sec:compatible}
To establish the compatibility between MFs computed on different maps, we follow an approach similar to the one adopted in \cite{PlanckVII20_isotropy}, where a thorough investigation of the statistical properties of CMB anisotropies was conducted on real \emph{Planck} data, comparing the MFs of a single component-separated map against those of a set of Gaussian simulations. Here we apply the same analysis, but on a simulated dataset (``data'' realisation) and repeated for $N = 300$ times. We will then provide an estimation of the MFs compatibility test over this sample of simulations. For the remainder of this Section, we will use ``Gaussian simulations'' to refer either to tensor (\texttt{CMBtens}) or Gaussian foreground residuals (\texttt{FGgauss}) simulations, as we carry out the compatibility analysis in the same way for the two sets.

MFs are evaluated for 12 thresholds $u$ between $-3$ and 3 in $\sigma$ units, with maps normalised by their variance. To quantify the level of deviation between MFs of different maps, $\chi^2$ values are computed assuming a Gaussian likelihood for the MFs at every threshold:
\begin{equation}
    \chi^2_s = \left[ y_s - \bar{y}_s^\text{sim} \right] ^\mathrm{T} \Sigma_s^{-1} \left[ y_s - \bar{y}_s^\text{sim} \right] , \qquad s=1, \, \dots, \, 300.
    \label{eq:chi2}
\end{equation}
where $y_s$ represents one of the three MFs ($y = v_0, v_1, v_2$) of the $s$-th \texttt{FGres} realisation, and $\bar{y}_s^\text{sim} \equiv \langle y_s^\text{sim} \rangle$ is the mean of the MFs computed on the 300 Gaussian simulations generated for that particular $s$-th ``data'' simulation. We recall that we have $N = 300$ realisations of \texttt{FGres} maps, and for each one of them we generate other 300 simulations of \texttt{CMBtens} or \texttt{FGgauss} maps. In the first comparison, MFs can detect deviations due to both the shape of the power spectrum and the non-Gaussianity, while in the second one the discrepancy is entirely caused by the presence of non-Gaussianity. \\
Correlations between different thresholds \cite{Ducout13} are taken into account using the appropriate covariance matrix $\Sigma_s$, different for each ``data'' realisation. An unbiased estimator of the inverse of the covariance is given by
\begin{equation}
    \Sigma_s^{-1} = \frac{N - d - 2}{N - 1} \mathbf{C}_s^{-1} ,
    \label{eq:cov_hart}
\end{equation}
where $N = 300$ is the number of simulations, $d = 12$ is the number of thresholds (degrees of freedom), and $\mathbf{C}_s$ is the sample covariance computed from the $N$ Gaussian simulations (tensor or foreground residuals):
\begin{equation}
    \mathbf{C} = C_{ij} \equiv \langle (y_i^\text{sim} - \bar{y}_i^\text{sim}) (y_j^\text{sim} - \bar{y}_j^\text{sim}) \rangle ,
    \label{eq:cov}
\end{equation}
with $i, j$ referring to MFs computed at different thresholds. The pre-factor in Eq.~\eqref{eq:cov_hart} is known as the Hartlap factor \cite{Hartlap07} and it is needed for the estimator to be unbiased. \\
For each ``data'' realisation ($\chi^2_\text{data} = \chi^2_s$), we compute its probability-to-exceed (PTE) which is the amount of the $\chi^2$ distribution of the MFs computed on Gaussian simulation to exceed the observed $\chi^2_\text{data}$ realisation:
\begin{equation}
    \text{PTE}_s = \chi^2_\text{sims} > \chi^2_\text{data}.
    \label{eq:pte}
\end{equation}
We assign a detection (i.e., MFs are not compatible) to the $s$-th simulation when $\text{PTE}_s < 0.05$, for at least one of the MFs. A detection means that statistically significant deviations are observed between MFs computed on a map contaminated by non-Gaussian residuals and the mean MFs of Gaussian realisations with similar power in tensor or foreground residuals. Obtaining many detections over a large number of simulations indicates that MFs are a reliable robustness test, while a small amount of detection would suggest that MFs are not able, most of the times, to distinguish between Gaussian and non-Gaussian contributions. \\
We define the test efficiency $\eta(A, B)$ between two sets of data $A$ and $B$ as the percentage of detection we get over all the simulations,
\begin{equation}
    \eta(A, B) = \frac{\# \left[ \, \text{detections} \, (A, B) \right]}{N} = \frac{\# \left[\text{PTE} < 0.05 \right]}{N} ,
    \label{eq:eta}
\end{equation}
where $\#[\cdot]$ stands for the number of simulations satisfying the condition in brackets. We will report this number to quantify the detection power of the MFs test, for each combination of component separation and foregrounds model.

\section{Results and discussion}
\label{sec:results}

\begin{figure}[ht]
    \centering
    \includegraphics[width=1.0\textwidth]{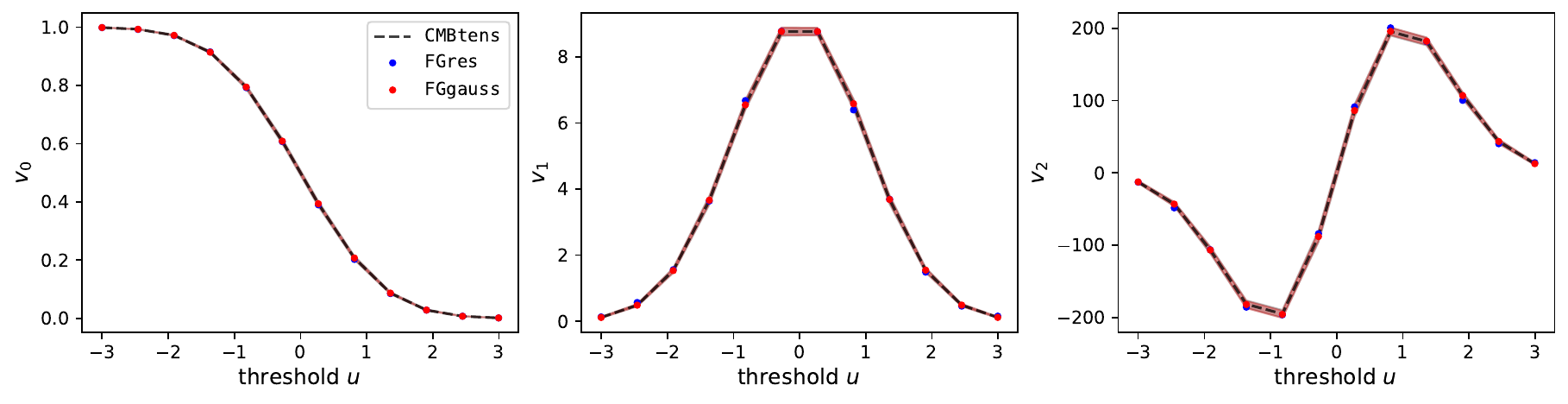}
    \includegraphics[width=1.0\textwidth]{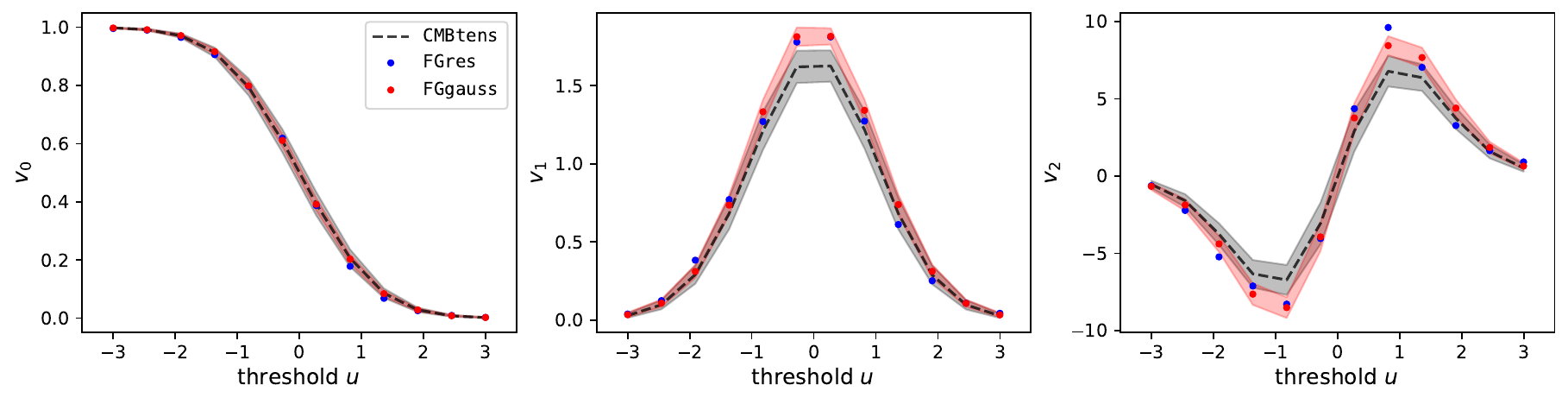}
    \caption{Minkowski functionals, for a single simulation, in the MC-NILC - \texttt{d10s5} case. MFs of the CMB solution (\texttt{FGres}) are shown with blue points; the black dashed line is the average MFs from 300 Gaussian realisations of tensors (\texttt{CMBtens}), and red points are the average MFs from 300 Gaussian realisations of foreground residuals (\texttt{FGgauss}). The correspondent gray and red shaded areas represent the standard deviation of the MFs across the simulations. \textit{Upper row}: real space (including all scales). \textit{Lower row}: needlet space (first band $b_0$, $\ell \lesssim 60$).}
    \label{fig:mfs}
\end{figure}

In this Section we report and discuss the results of this work. Tables~\ref{tab:results_pixel} and \ref{tab:results_needlet} offer a comprehensive report of the results for the scenarios considered in the analysis, while Figure~\ref{fig:mfs} shows an example of MFs comparison for the MC-NILC - \texttt{d10s5} case as we consider it as our most realistic scenario, since \texttt{d10s5} is the model closer to observations and MC-NILC is the state-of-the-art minimum-variance component separation algorithm which will be used in future CMB experiments. In the Appendix~\ref{sec:appendix} we report examples for the NILC - \texttt{d1s1} scenario, which represents a case with more significant foreground residuals on the large scales.

We report our results in terms of the compatibility level between MFs of different maps. The efficiency of the test (percentage of incompatible simulations, Eq.~\eqref{eq:eta}) is the main summary statistic of the study, reported for all the combinations of component separation technique and foregrounds modelling, in pixel and needlet space. $\eta(\texttt{FGres}, \texttt{CMBtens})$ indicates the values computed by comparing MFs against those of Gaussian simulations of tensor perturbations, while $\eta(\texttt{FGres}, \texttt{FGgauss})$ refers to the comparison with Gaussian simulations of foreground residuals (both co-added with CMB lensing and noise residuals).

\begin{figure}
    \centering
    \includegraphics[width=1.0\textwidth]{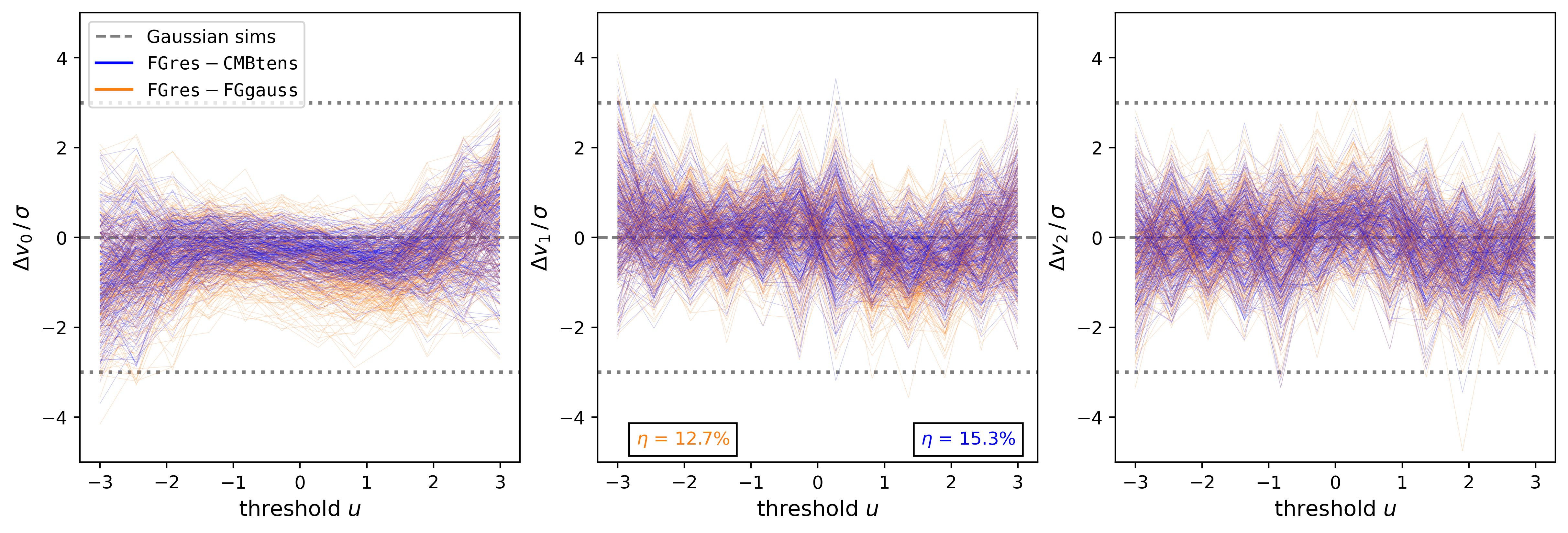}
    \includegraphics[width=1.0\textwidth]{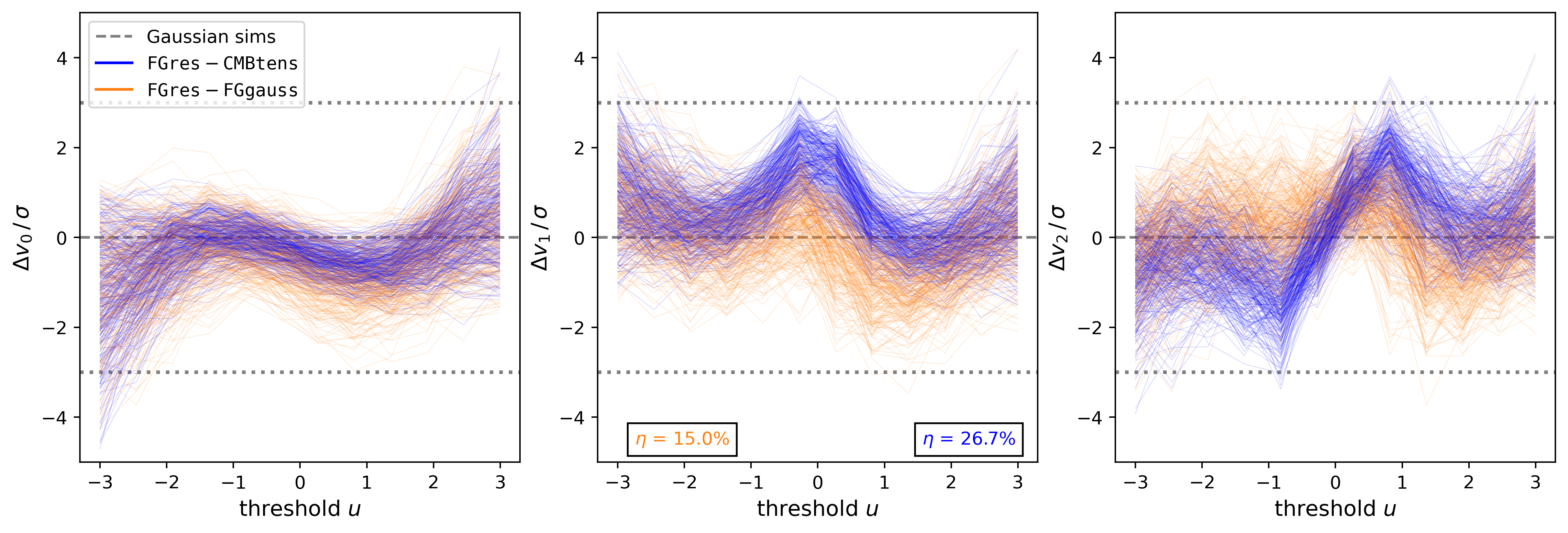}
    \caption{MFs weighted difference for the MC-NILC - \texttt{d10s5} scenario, referring to \texttt{FGres} against \texttt{CMBtens} maps (blue lines) and \texttt{FGres} against \texttt{FGgauss} maps (orange lines). See text for details. Plots are shown for pixel (upper row) and needlet (band $b_0$, lower row) analysis.}
    \label{fig:mfs_residuals}
\end{figure}

Figure~\ref{fig:mfs} shows an example for one simulation of the computed MFs in pixel and needlet space (first band $b_0$), reporting MFs of the single CMB solution (blue points) along with the average ones of the \texttt{CMBtens} (dashed black line) and \texttt{FGgauss} (red points) simulations. We also show, in Figure~\ref{fig:mfs_residuals} and for all the 300 simulations, the MFs relative difference in units of the standard deviation of the simulations $\Delta v / \sigma = \left( y - \bar{y}^\text{sim} \right) / \sigma$; each line corresponds to the difference between one CMB solution realisation and the means of the two sets of Gaussian simulations, blue for \texttt{CMBtens} and orange for \texttt{FGgauss} simulations.

\begin{table}
    \centering
    \caption{Efficiency $\eta$ (in \%) of the MFs test on polarisation $B$-mode maps, in pixel space (i.e., considering all scales at the same time), both for no delensing ($A_\text{lens} = 1.0$) or 50\% delensing ($A_\text{lens} = 0.5$) applied. Values are reported for each scenario considered in the analysis. $\eta(A, B)$ indicates the efficiency computed by comparing the dataset $A$ against the reference dataset $B$.}
    \begin{tabular}{c c m{1.5cm} m{1.5cm} m{1.5cm}}
        \\
        \multicolumn{5}{c}{no delensing ($A_\text{lens} = 1$)} \\
        \hline
        & $A, B$ & \texttt{d0s0} & \texttt{d1s1} & \texttt{d10s5} \\ \hline
        \multirow{2}{2cm}{NILC} & \texttt{FGres}, \texttt{CMBtens} & 12.7 & 95.3 & 88.3 \\ [0.25ex]
        & \texttt{FGres}, \texttt{FGgauss} & 11.0 & 39.3 & 26.7 \\ \hline
        \multirow{2}{2cm}{MC-NILC} & \texttt{FGres}, \texttt{CMBtens} &  & 13.3 & 15.3 \\ [0.25ex]
        & \texttt{FGres}, \texttt{FGgauss} &  & 10.3 & 12.7 \\
        \hline
        \\
        \multicolumn{5}{c}{50\% delensing ($A_\text{lens} = 0.50$)} \\
        \hline
        & $A, B$ & \texttt{d0s0} & \texttt{d1s1} & \texttt{d10s5} \\ \hline
        \multirow{2}{2cm}{NILC} & \texttt{FGres}, \texttt{CMBtens} & 10.3 & 100 & 95.3 \\ [0.25ex]
        & \texttt{FGres}, \texttt{FGgauss} & 11.3 & 42.7 & 27.7 \\ \hline
        \multirow{2}{2cm}{MC-NILC} & \texttt{FGres}, \texttt{CMBtens} &  & 14.0 & 20.3 \\ [0.25ex]
        & \texttt{FGres}, \texttt{FGgauss} &  & 15.0 & 24.0 \\
        \hline
    \end{tabular}
    \label{tab:results_pixel}
\end{table}

\begin{table}
    \centering
    \caption{Same as Table~\ref{tab:results_pixel}, but in needlet space, for the first band $b_0$ (largest scales, $\ell \lesssim 60$).}
    \begin{tabular}{c c m{1.5cm} m{1.5cm} m{1.5cm}}
        \\
        \multicolumn{5}{c}{no delensing ($A_\text{lens} = 1$)} \\
        \hline
        & $A, B$ & \texttt{d0s0} & \texttt{d1s1} & \texttt{d10s5} \\ \hline
        \multirow{2}{2cm}{NILC} & \texttt{FGres}, \texttt{CMBtens} & 10.7 & 93.0 & 51.7 \\ [0.25ex]
        & \texttt{FGres}, \texttt{FGgauss} & 11.7 & 72.7 & 23.3 \\ \hline
        \multirow{2}{2cm}{MC-NILC} & \texttt{FGres}, \texttt{CMBtens} &  & 22.3 & 26.7 \\ [0.25ex]
        & \texttt{FGres}, \texttt{FGgauss} &  & 16.0 & 15.0 \\
        \hline
        \\
        \multicolumn{5}{c}{50\% delensing ($A_\text{lens} = 0.50$)} \\
        \hline
        & $A, B$ & \texttt{d0s0} & \texttt{d1s1} & \texttt{d10s5} \\ \hline
        \multirow{2}{2cm}{NILC} & \texttt{FGres}, \texttt{CMBtens} & 11.7 & 95.7 & 57.0 \\ [0.25ex]
        & \texttt{FGres}, \texttt{FGgauss} & 11.0 & 75.3 & 20.0 \\ \hline
        \multirow{2}{2cm}{MC-NILC} & \texttt{FGres}, \texttt{CMBtens} &  & 24.3 & 42.0 \\ [0.25ex]
        & \texttt{FGres}, \texttt{FGgauss} &  & 18.3 & 18.0 \\
        \hline
    \end{tabular}
    \label{tab:results_needlet}
\end{table}

Concerning the real-space analysis, we find a high efficiency of the test when comparing MFs against Gaussian tensor simulations, while using NILC on the \texttt{d1s1} and \texttt{d10s5} models, with $\eta(\texttt{FGres}, \texttt{CMBtens}) \simeq 95\%$ and 88\% respectively. This means that over 300 simulations, MFs are able to detect the discrepancy due to foregrounds contamination in 95\% (or 88\%) of the cases. However, this deviation is mainly driven by the large differences at the power spectrum level between the two maps: as discussed in Section~\ref{sec:mfs}, the MFs expected values depend through Eq.~\eqref{eq:gkf} both on the level of Gaussianity in the map, and on the shape of the power spectrum (included in the $\mu$ term). MFs computed on maps with mismatching power spectrum features will deviate between each other, with discrepancies sourced by both contributions. As already pointed out in Section~\ref{sec:sims_gauss_fgs}, the large-scale shapes of the power spectra do not match, for the two scenarios discussed in this paragraph (for comparison, the percentages of simulations providing a bad fit are $\sim 98\%$ for \texttt{d1s1} and 74\% for \texttt{d10s5}). Hence, one could already rule out the cosmological origin of the signal by evaluating the goodness-of-fit, and the hypothesis would then be confirmed by the MFs test; we notice that in some cases, such as NILC - \texttt{d10s5}, we find that the MFs compatibility test is more powerful than the power spectrum goodness of fit.

The same conclusion can also be drawn from the lower value of the efficiency when Gaussian foreground residuals simulations are considered. We find $\eta(\texttt{FGres}, \texttt{FGgauss}) \simeq 39\%$ and $\simeq 27\%$ for \texttt{d1s1} and \texttt{d10s5}, respectively, when NILC is applied. This amount of detection is entirely due to the different statistics in one component of the two datasets, since the corresponding power spectra are equivalent. This interpretation of the results is also confirmed by the detection efficiencies obtained when applying MC-NILC, where the large-scale features of the power spectra are more similar: the number of detections is relatively low, and close to what we find in the \texttt{d0s0} scenario, where the tensor-to-scalar ratio is not biased and for which we do not expect a statistically significant detection.

In the needlet-based analysis, we compute MFs on filtered maps adopting three needlet bands which sample separate range of multipoles; here we focus on the first band $b_0$ which is the most relevant for our analysis ($\ell \lesssim 60$), as we expect the most relevant sources of bias to the tensor-to-scalar ratio to be in this multipole range. In this way, different contributions from different sky components should be isolated, in particular on the largest scales where foreground residuals dominate over the CMB and noise residuals components; the MFs detection power is expected to be boosted in the first needlet band as the higher multipoles noise contribution is filtered out, thus removing a Gaussian component at the map level and enhancing non-Gaussian features typical of foreground residuals. In general, we observe a slight increase in the detection number for some of the cases (e.g., for MC-NILC), while in others the efficiency remains the same or is even reduced. This is due to the fact that the tensor-to-scalar ratio used for generating the simulations is fitted by considering only the multipoles included in the needlet band, providing a better fit to the power spectrum and thus reducing deviations arising from the different spectral shapes. The detection number still remains considerably low and close to the ``no detection'' level typical of the \texttt{d0s0} case.

We can explain why MFs can not clearly detect non-Gaussianity by looking at the maps shown in Figure~\ref{fig:mfs_maps_pixel} and \ref{fig:mfs_maps_needlet}: the Gaussian component (due to CMB lensing and noise residuals) is dominating the structures of the maps, hiding the lower contribution coming from non-Gaussian foreground residuals. This suggests that the level of foreground residuals is too low with respect to the noise residuals in order to be detected by MFs. Indeed, in the NILC - \texttt{d1s1} case foreground residuals are relatively larger than noise residuals with respect to other scenarios, and we register a higher number of detections.

This can be straightforwardly tested by repeating the analysis after removing CMB and noise contributions, looking at discrepancies between the non-Gaussian (lower left map of Figure~\ref{fig:mfs_maps_pixel}) and Gaussian (lower middle map) foreground residuals; in this case, the efficiency of the MFs test is always 100\%, meaning that MFs are perfectly able to distinguish the non-Gaussianity. Once added, CMB lensing and noise residuals become the driving contribution in the MFs and tend to overwhelm the foregrounds, hindering the MFs detection power.

To better understand the relative contribution of CMB lensing and noise residuals, we repeat the same analysis while applying two independent delensing and denoising factors on the CMB and noise residuals components, respectively, after the component separation. In this way, we are removing part of the Gaussian contribution at the map level, in order to check which is the dominant component and its impact on the ability of MFs to detect non-Gaussianity. In Tables~\ref{tab:results_pixel} and \ref{tab:results_needlet} we report results for a realistically attainable delensing of $50\%$ ($A_\text{lens} = 0.5$), showing that in general we obtain an increase of the efficiency $\eta$ as expected: we reduced one of the Gaussian contribution, and the anisotropy induced by foreground residuals is more evident to MFs. However, the detection number in the most realistic scenario (MC-NILC - \texttt{d10s5}) is still considerably low, being $<30\%$. For this particular case only, we also rescale the noise residuals maps by a denoising factor of $75\%$ ($A_\text{noise} = 0.25$), observing an efficiency raise to $93\%$ and $92\%$ for MFs compared with \texttt{CMBtens} and \texttt{FGgauss} maps, respectively, in pixel space. In the needlet-based analysis, the efficiencies reach $\sim 99\%$ in both cases. This confirms the role of noise residuals as the dominant component over CMB lensing, driving the Gaussian contribution. It also provides the noise level needed for the MFs to be considered as reliable tools for this robustness test.

It is also worth to discuss the \texttt{d10s5} scenario, which at first sight provides a counter-intuitive result: in general we expect that an increase in foregrounds complexity should correspond to an increase in the number of detected deviations, with respect to the simpler \texttt{d1s1} case, as the foreground residuals should be more significant. Instead, we obtain a lower amount of detections for \texttt{d10s5}, both in pixel- and needlet-based analysis. This can be explained by considering that in general, the application of a minimum-variance component separation pipeline to a more complex foreground model results in the increase of both foreground and noise residuals. It means that the CMB solution will contain more contamination due to foreground residuals, but it will also carry a stronger Gaussian component driven by noise residuals, which is responsible for reducing the MFs detection power (as discussed in the previous paragraph). An additional analysis on this is reported in the Appendix~\ref{sec:appendix}.

Following the results of the comparison between MFs of \texttt{FGres} and \texttt{FGgauss} maps, we conclude that, with the obtained relative amount of foregrounds and noise residuals in the component separated maps, MFs will not be able to detect the non-cosmological nature of a potentially biased measurement of the tensor-to-scalar ratio; the level of noise residuals is masking the anisotropic non-Gaussian contribution due to foreground residuals, preventing MFs from correctly identifying the contamination. The false cosmological signal passes the MFs test in the majority of the simulations, while not providing additional information on the astrophysical nature of the contamination. Thus, MFs can not be reliably used as higher-order statistical tool for a robustness test in the event of a future tensor-to-scalar ratio detection with a \emph{LiteBIRD}-like configuration. However, they may still be useful in scenarios where foreground residuals are relatively more dominant than noise residuals.

\section{Conclusions}
\label{sec:conclusions}
Future CMB experiments like the \emph{LiteBIRD} satellite will target a detection of the primordial tensor perturbations with an overall uncertainty on the tensor-to-scalar ratio of $\delta(r=0) = 0.001$. The presence of foreground residuals left after the application of component separation algorithms could bias this measurement, providing a false detection of the primordial signal.

In this work, we explored the possibility of using Minkowski functionals (MFs) to validate an eventual detection of the tensor-to-scalar ratio, exploiting the sensitivity of this higher-order statistics to the presence of non-Gaussianity induced by foreground residuals. We applied two minimum-variance component separation algorithms (NILC and MC-NILC) to realistic sky simulation of CMB, noise, and three different models of foregrounds (\texttt{d0s0}, \texttt{d1s1}, \texttt{d10s5}) in order to obtain cleaned $B$-mode maps. After evaluating the likelihood function on their angular power spectrum, we recover a (biased) tensor-to-scalar ratio for each realisation. We used these inferred values of $r$ to generate Gaussian simulations of tensor perturbations, and then we compared the MFs of these simulations with the ones evaluated on the CMB solution (the output map of the component separation), in order to check their compatibility and establish if the MFs can spot the discrepancy between the two sets of maps due to the foregrounds non-Gaussianity. Since MFs expected values depend both on the shape of the power spectrum and statistics of the considered maps, we also compare the CMB solution MFs with those computed on a set of maps that have the same CMB lensing and noise residuals, but with the non-Gaussian foreground residuals replaced by a Gaussian counterpart with the same power spectrum. In this way, the eventual deviation detected by MFs is entirely driven by non-Gaussianity.

The general outcome of the study is that MFs, with the obtained relative level of foreground and noise residuals for the considered instrumental configuration and component separation techniques, are not able to significantly detect the presence of non-Gaussianity in component-separated maps, even if this contamination is biasing the value of the inferred tensor-to-scalar ratio. This result has been observed both when using all the scales for MFs computation, or considering needlet-filtered maps (isolating the largest scales). In the majority of the cases, the test is passed and no warning is raised, meaning that in real life we would accept a false signal as a real measurement of $r$. This means that MFs can not be used as a reliable robustness test in the context of a map-based inference of the tensor-to-scalar ratio, at least for a \emph{LiteBIRD}-like experiment which adopts MC-NILC as component separation algorithm. An analogous analysis could be applied to other experimental configurations, for instance to a ground-based CMB telescope like SO \cite{SO19}, where only a partial region of the sky is observed, and for which the MFs study could provide different results.

MFs are not the only higher-order statistical tools available for CMB studies. For instance, wavelet scattering transform (WST) \cite{Allys19,Blancard20,Delouis22,Mousset2024,Campeti2025} are low-variance statistical descriptor of non-Gaussian processes introduced in the field of data science, able to provide summary statistics of CMB polarisation maps. Also the bispectrum estimators \cite{PlanckIX18_ng} could represent a powerful tool for the study on non-Gaussian features in component-separated CMB maps. We will address the exploration of such tools in the context of robustness test for tensor-to-scalar measurements in a future work.

\acknowledgments

We thank Anthony Banday, Luca Pagano, and Giacomo Galloni for the useful comments, feedbacks, and discussions in the development of this work. The authors acknowledge partial support by the Italian Space Agency (ASI Grants No. 2020-9-HH.0 and 2016-24-H.1-2018) and the Euclid Project, as well as the RadioForegroundsPlus Project HORIZON-CL4-2023-SPACE-01, GA 101135036 and through the Project SPACE-IT-UP by the Italian Space Agency and Ministry of University and Research, Contract Number 2024-5-E.0. The authors also acknowledge partial support by the InDark and LiteBIRD Initiative of the National Institute for Nuclear Physics.

\appendix
\section{Appendix}
\label{sec:appendix}

\subsection{Goodness-of-fit analysis}
\label{sec:gof}
\begin{figure}
    \centering
    \includegraphics[width=1.\textwidth]{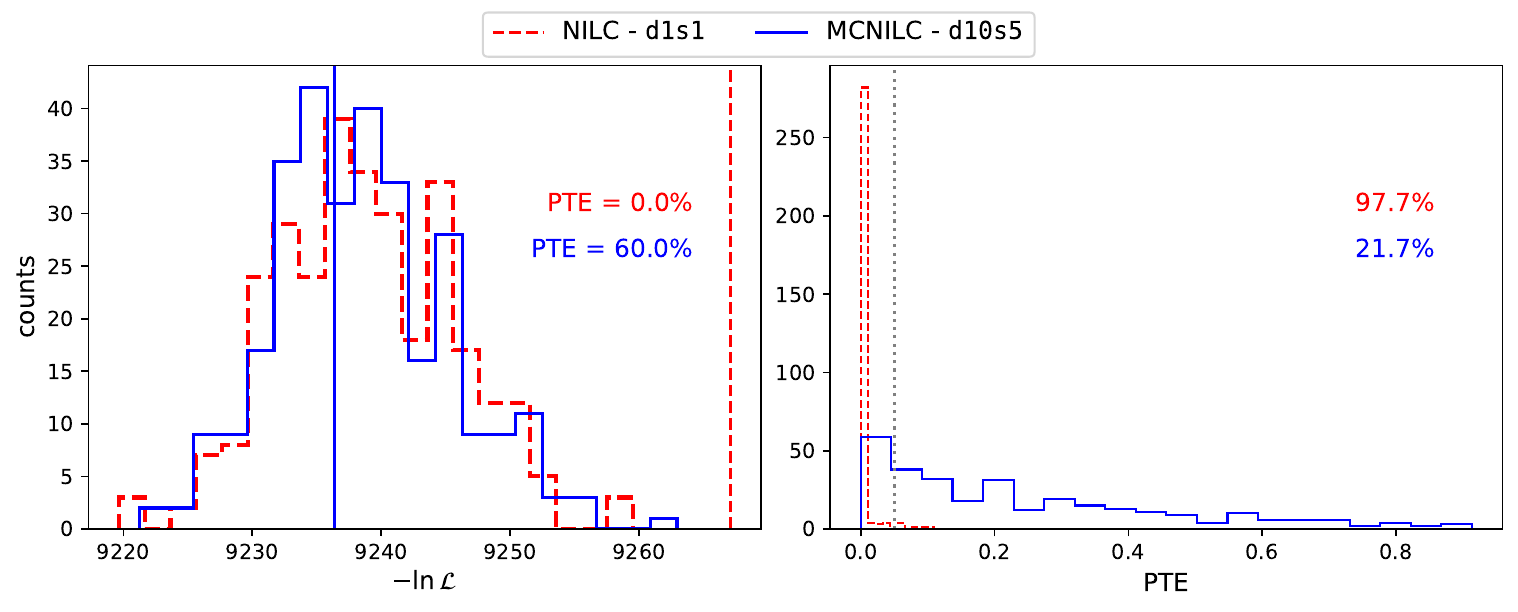}
    \caption{Results of the goodness-of-fit analysis for the NILC - \texttt{d1s1} (red dashed lines) and MC-NILC - \texttt{d10s5} (blue solid lines) cases. The left panel shows one realisation of the $- \ln \mathcal{L}$ distribution of the 300 tensor simulations, compared against the value given by the best fit on the CMB solution (vertical lines); the corresponding PTE is also indicated. The right panel reports the PTE distribution for the 300 simulations, with the percentage of incompatible simulations shown in the upper right.}
    \label{fig:logL_ptes}
\end{figure}

In this Section we show some examples of the goodness-of-fit analysis carried out during the inference of the tensor-to-scalar ratio (described in Section~\ref{sec:likelihood}). Figure~\ref{fig:logL_ptes} reports results for NILC - \texttt{d1s1} and MC-NILC \texttt{d10s5} scenarios, with the $-\ln{\mathcal{L}}$ distribution for one simulation on the left, and the PTE distribution on the right, along with the percentage of simulations failing the $\chi^2$ test. For this particular realisation, the NILC - \texttt{d1s1} case fails the test ($\text{PTE}<0.05$) while for MC-NILC - \texttt{d10s5} the test is passed ($\text{PTE}>0.05$).

\subsection{Examples of MFs in other scenarios}
\label{sec:other_mfs}
\begin{figure}
    \centering
    \includegraphics[width=1.0\textwidth]{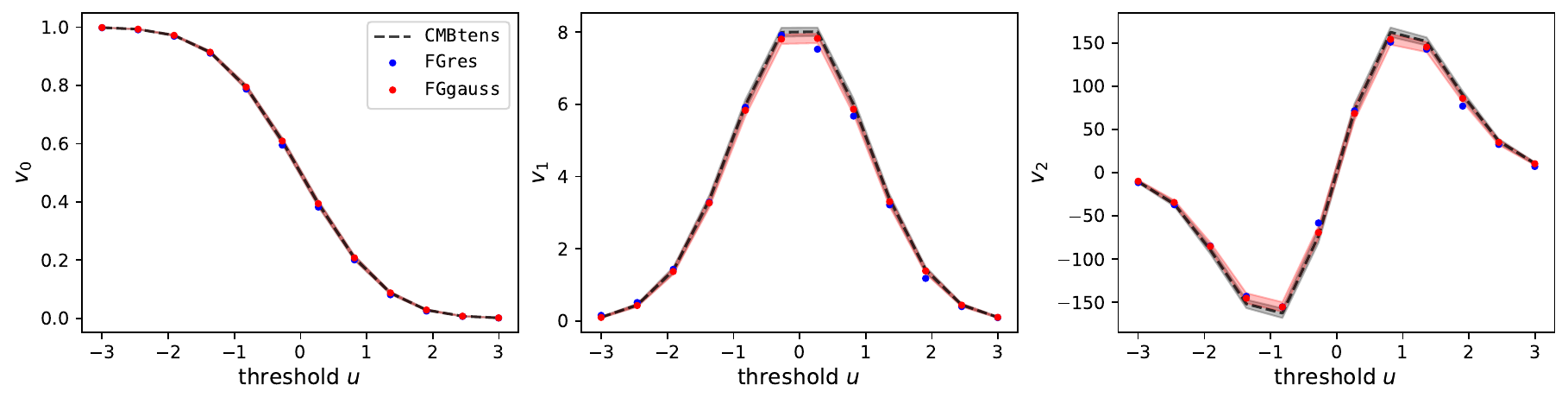}
    \includegraphics[width=1.0\textwidth]{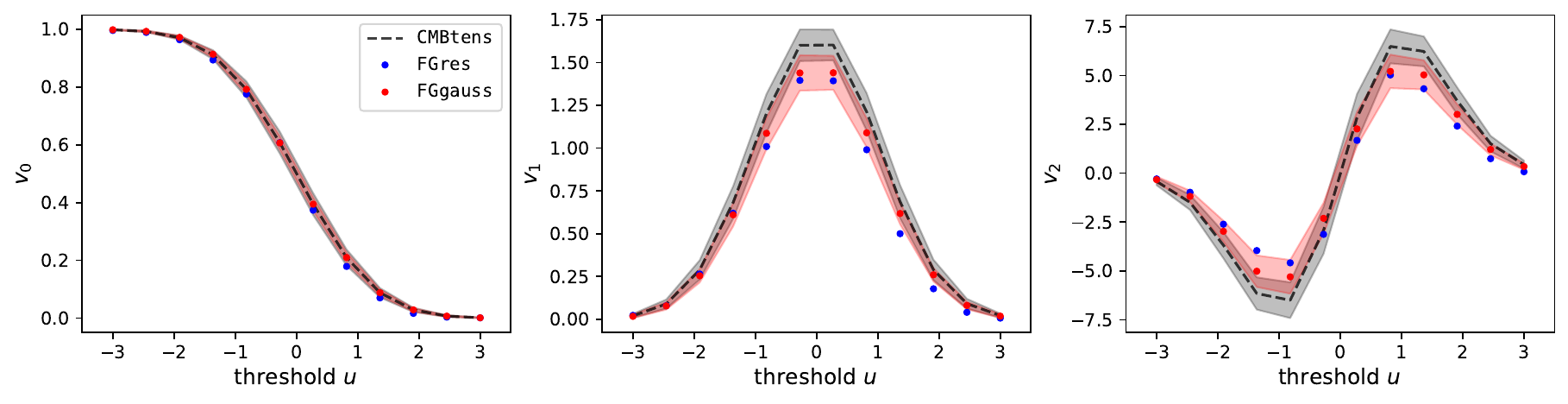}
    \caption{Same as Figure~\ref{fig:mfs}, but for the NILC - \texttt{d1s1} scenario. \textit{Upper row}: pixel space (including all scales). \textit{Lower row}: needlet space (first band, $\ell \lesssim 60$).}
    \label{fig:mfs_nilc}
\end{figure}

\begin{figure}
    \centering
    \includegraphics[width=1.\textwidth]{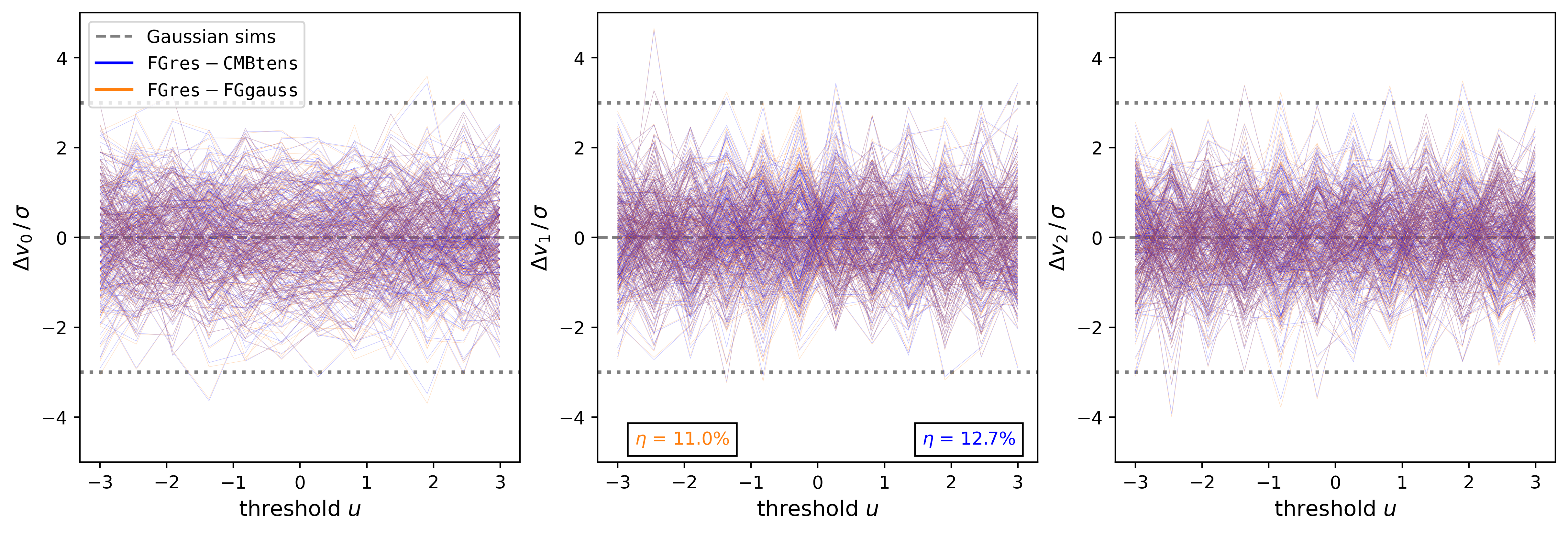}
    \includegraphics[width=1.\textwidth]{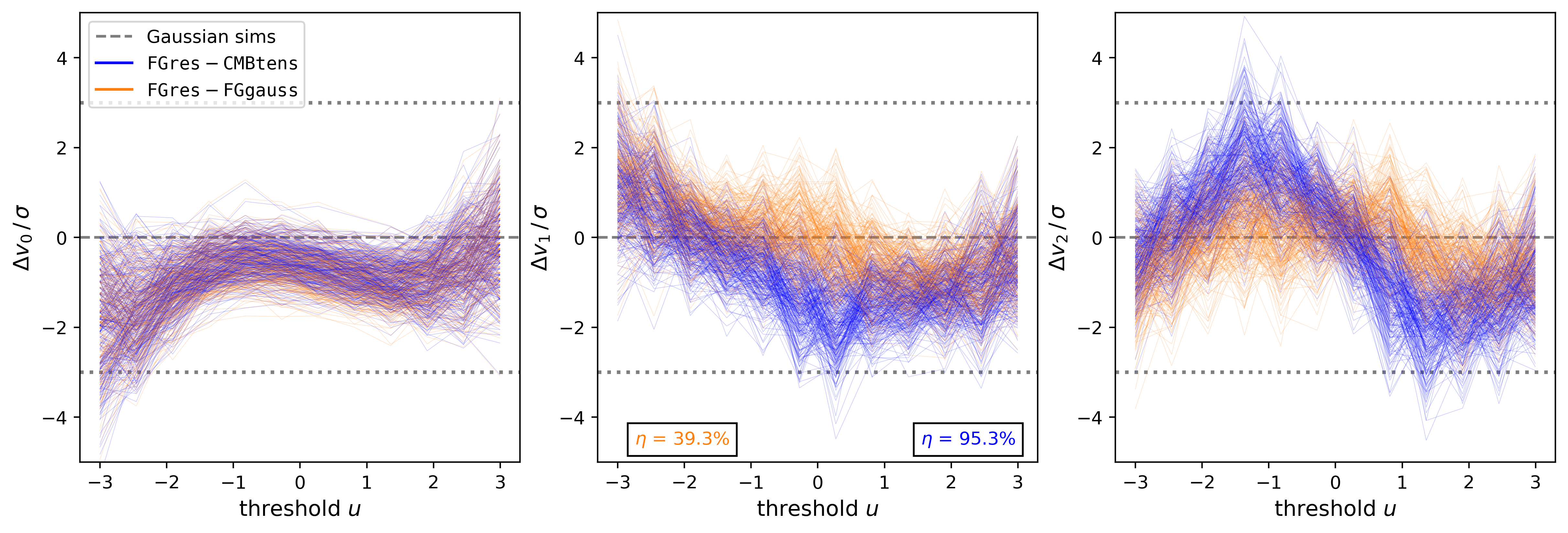}
    \caption{Same as Figure~\ref{fig:mfs_residuals}, but for the NILC - \texttt{d0s0} (upper row) and NILC - \texttt{d1s1} (lower row) cases, in pixel space (including all scales).}
    \label{fig:mfs_residuals_nilc}
\end{figure}

Here we report MFs evaluated in other scenarios considered in this work. In Figure~\ref{fig:mfs_nilc} we illustrate the NILC - \texttt{d1s1} case both in pixel and needlet space (largest scales), while in Figure~\ref{fig:mfs_residuals_nilc} we show the MFs weighted difference as presented in Figure~\ref{fig:mfs_residuals}, for the NILC algorithm applied on the \texttt{d0s0} and \texttt{d1s1} foreground models. The \texttt{d0s0} panels are useful to visualize a configuration where we have no pattern in the distribution of simulations, compatible with noise-like features. See the discussion in Section~\ref{sec:results} for more details.

\subsection{Dependence of MFs test efficiency on the foreground model complexity}
\begin{figure}
    \centering
    \includegraphics[width=1.\textwidth]{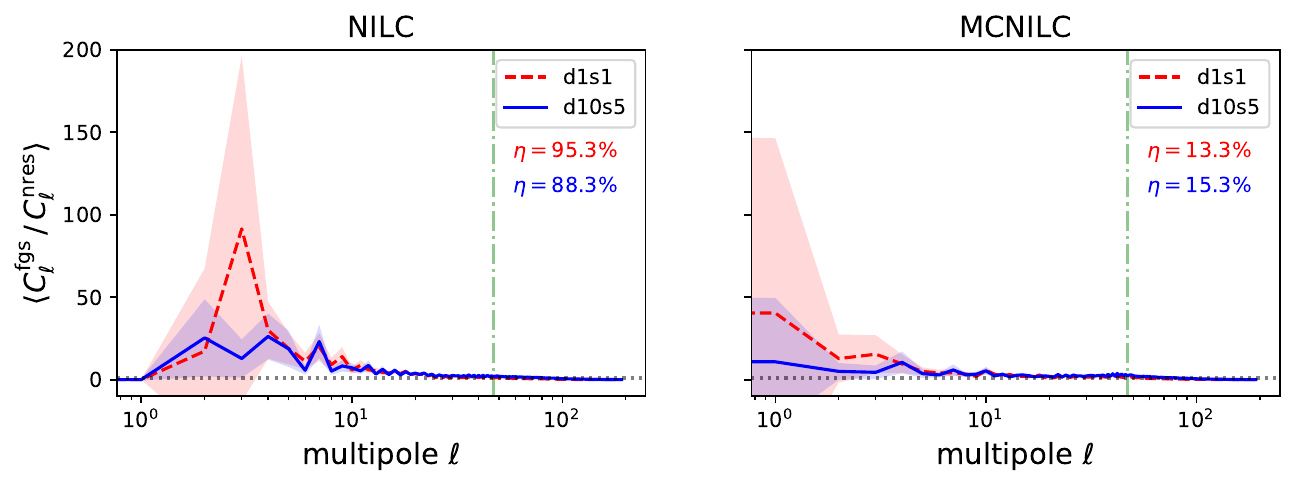}
    \caption{Average ratio between spectra of foreground and noise residuals. Dashed red lines indicate the \texttt{d1s1} model, while blue solid lines are for \texttt{d10s5}. Shaded areas represent the standard deviation across 300 simulations. MFs test efficiencies are also reported for comparison. The vertical line indicates the smallest scale included in the first needlet band $b_0$.}
    \label{fig:res_ratio}
\end{figure}

In Section~\ref{sec:results} we found a decrease in the MFs efficiency when moving from a lower (\texttt{d1s1}) to a higher (\texttt{d10s5}) foreground complexity, and we addressed it to the higher level of noise residuals present in the latter case. To explore this hypothesis, for each simulation we evaluate the relative amount of residuals by computing the ratio between the power spectra of foreground ($C_\ell^\text{fgs}$) and noise residuals ($C_\ell^\text{nres}$). In Figure~\ref{fig:res_ratio} we report the mean of this quantity across simulations: for the NILC algorithm (left panel) the average ratio in the \texttt{d10s5} case is slightly lower than \texttt{d1s1}, meaning that the separation between foregrounds and noise is less clear for MFs to detect, thus leading to a lower efficiency $\eta$. For the MC-NILC case the residual ratio is similar for the two models, leading to roughly the same number of detections. These considerations suggest that the MFs efficiency is not heavily dependent on the complexity of the foreground model, but rather on the relative levels of foregrounds and noise residuals, which can both increase or decrease based on the performance of the component separation.

\bibliographystyle{JHEP}
\bibliography{mfs.bib}

\providecommand{\href}[2]{#2}\begingroup\raggedright\begin{thebibliography}{10}

\bibitem{WMAP13_results}
C.L.~{Bennett}, D.~{Larson}, J.L.~{Weiland}, N.~{Jarosik}, G.~{Hinshaw}, N.~{Odegard} et~al., \emph{{Nine-year Wilkinson Microwave Anisotropy Probe (WMAP) Observations: Final Maps and Results}}, \href{https://doi.org/10.1088/0067-0049/208/2/20}{\emph{\apjs} {\bfseries 208} (2013) 20} [\href{https://arxiv.org/abs/1212.5225}{{\ttfamily 1212.5225}}].

\bibitem{PlanckVI20_params}
{Planck Collaboration}, N.~{Aghanim}, Y.~{Akrami}, M.~{Ashdown}, J.~{Aumont}, C.~{Baccigalupi} et~al., \emph{{Planck 2018 results. VI. Cosmological parameters}}, \href{https://doi.org/10.1051/0004-6361/201833910}{\emph{\aap} {\bfseries 641} (2020) A6} [\href{https://arxiv.org/abs/1807.06209}{{\ttfamily 1807.06209}}].

\bibitem{ACT_DR4_20}
S.~{Aiola}, E.~{Calabrese}, L.~{Maurin}, S.~{Naess}, B.L.~{Schmitt}, M.H.~{Abitbol} et~al., \emph{{The Atacama Cosmology Telescope: DR4 maps and cosmological parameters}}, \href{https://doi.org/10.1088/1475-7516/2020/12/047}{\emph{\jcap} {\bfseries 2020} (2020) 047} [\href{https://arxiv.org/abs/2007.07288}{{\ttfamily 2007.07288}}].

\bibitem{Brout78}
R.~Brout, F.~Englert and E.~Gunzig, \emph{{The creation of the universe as a quantum phenomenon}}, \href{https://doi.org/https://doi.org/10.1016/0003-4916(78)90176-8}{\emph{Annals of Physics} {\bfseries 115} (1978) 78}.

\bibitem{Starobinsky80}
A.~Starobinsky, \emph{{A new type of isotropic cosmological models without singularity}}, \href{https://doi.org/https://doi.org/10.1016/0370-2693(80)90670-X}{\emph{Physics Letters B} {\bfseries 91} (1980) 99}.

\bibitem{Guth81}
A.H.~Guth, \emph{{Inflationary universe: A possible solution to the horizon and flatness problems}}, \href{https://doi.org/10.1103/PhysRevD.23.347}{\emph{Phys. Rev. D} {\bfseries 23} (1981) 347}.

\bibitem{Kamionkowski97}
M.~Kamionkowski, A.~Kosowsky and A.~Stebbins, \emph{{A Probe of Primordial Gravity Waves and Vorticity}}, \href{https://doi.org/10.1103/PhysRevLett.78.2058}{\emph{Phys. Rev. Lett.} {\bfseries 78} (1997) 2058}.

\bibitem{Hu97}
W.~Hu and M.~White, \emph{{A CMB polarization primer}}, \href{https://doi.org/https://doi.org/10.1016/S1384-1076(97)00022-5}{\emph{New Astronomy} {\bfseries 2} (1997) 323}.

\bibitem{Seljak_Zaldarriaga97_gw}
U.~Seljak and M.~Zaldarriaga, \emph{{Signature of Gravity Waves in the Polarization of the Microwave Background}}, \href{https://doi.org/10.1103/PhysRevLett.78.2054}{\emph{Phys. Rev. Lett.} {\bfseries 78} (1997) 2054}.

\bibitem{Planck20_inflation}
{Planck Collaboration}, {Akrami, Y.}, {Arroja, F.}, {Ashdown, M.}, {Aumont, J.}, {Baccigalupi, C.} et~al., \emph{{Planck 2018 results - X. Constraints on inflation}}, \href{https://doi.org/10.1051/0004-6361/201833887}{\emph{A\&A} {\bfseries 641} (2020) A10}.

\bibitem{Tristram22}
M.~Tristram, A.J.~Banday, K.M.~G\'orski, R.~Keskitalo, C.R.~Lawrence, K.J.~Andersen et~al., \emph{{Improved limits on the tensor-to-scalar ratio using BICEP and $Planck$ data}}, \href{https://doi.org/10.1103/PhysRevD.105.083524}{\emph{Phys. Rev. D} {\bfseries 105} (2022) 083524}.

\bibitem{Galloni23}
{Galloni, G. and Bartolo, N. and Matarrese, S. and Migliaccio, M. and Ricciardone, A. and Vittorio, N.}, \emph{Updated constraints on amplitude and tilt of the tensor primordial spectrum}, \href{https://doi.org/10.1088/1475-7516/2023/04/062}{\emph{Journal of Cosmology and Astroparticle Physics} {\bfseries 2023} (2023) 062}.

\bibitem{Kamionkowski_Kovetz16}
M.~{Kamionkowski} and E.D.~{Kovetz}, \emph{{The Quest for B Modes from Inflationary Gravitational Waves}}, \href{https://doi.org/10.1146/annurev-astro-081915-023433}{\emph{\araa} {\bfseries 54} (2016) 227} [\href{https://arxiv.org/abs/1510.06042}{{\ttfamily 1510.06042}}].

\bibitem{LiteBIRD23PTEP}
{LiteBIRD Collaboration}, E.~{Allys}, K.~{Arnold}, J.~{Aumont}, R.~{Aurlien}, S.~{Azzoni} et~al., \emph{{Probing cosmic inflation with the LiteBIRD cosmic microwave background polarization survey}}, \href{https://doi.org/10.1093/ptep/ptac150}{\emph{Progress of Theoretical and Experimental Physics} {\bfseries 2023} (2023) 042F01} [\href{https://arxiv.org/abs/2202.02773}{{\ttfamily 2202.02773}}].

\bibitem{BICEP21}
P.~{Ade}, Z.~{Ahmed}, M.~{Amiri}, D.~{Barkats}, R.B.~{Thakur}, C.A.~{Bischoff} et~al., \emph{{Improved Constraints on Primordial Gravitational Waves using Planck, WMAP, and BICEP/Keck Observations through the 2018 Observing Season}}, \href{https://doi.org/10.1103/PhysRevLett.127.151301}{\emph{\prl} {\bfseries 127} (2021) 151301} [\href{https://arxiv.org/abs/2110.00483}{{\ttfamily 2110.00483}}].

\bibitem{SO19}
P.~Ade, J.~Aguirre, Z.~Ahmed, S.~Aiola, A.~Ali, D.~Alonso et~al., \emph{{The Simons Observatory: science goals and forecasts}}, \href{https://doi.org/10.1088/1475-7516/2019/02/056}{\emph{Journal of Cosmology and Astroparticle Physics} {\bfseries 2019} (2019) 056}.

\bibitem{S419}
K.~{Abazajian}, G.~{Addison}, P.~{Adshead}, Z.~{Ahmed}, S.W.~{Allen}, D.~{Alonso} et~al., \emph{{CMB-S4 Science Case, Reference Design, and Project Plan}}, \href{https://doi.org/10.48550/arXiv.1907.04473}{\emph{arXiv e-prints} (2019) arXiv:1907.04473} [\href{https://arxiv.org/abs/1907.04473}{{\ttfamily 1907.04473}}].

\bibitem{Lewis06_lensing}
A.~{Lewis} and A.~{Challinor}, \emph{{Weak gravitational lensing of the CMB}}, \href{https://doi.org/10.1016/j.physrep.2006.03.002}{\emph{\physrep} {\bfseries 429} (2006) 1} [\href{https://arxiv.org/abs/astro-ph/0601594}{{\ttfamily astro-ph/0601594}}].

\bibitem{Zaldarriaga_Seljak98_lensing}
M.~Zaldarriaga and U.~Seljak, \emph{{Gravitational lensing effect on cosmic microwave background polarization}}, \href{https://doi.org/10.1103/PhysRevD.58.023003}{\emph{Phys. Rev. D} {\bfseries 58} (1998) 023003}.

\bibitem{SPT15}
E.J.~{Baxter}, R.~{Keisler}, S.~{Dodelson}, K.A.~{Aird}, S.W.~{Allen}, M.L.N.~{Ashby} et~al., \emph{{A Measurement of Gravitational Lensing of the Cosmic Microwave Background by Galaxy Clusters Using Data from the South Pole Telescope}}, \href{https://doi.org/10.1088/0004-637X/806/2/247}{\emph{\apj} {\bfseries 806} (2015) 247} [\href{https://arxiv.org/abs/1412.7521}{{\ttfamily 1412.7521}}].

\bibitem{ACT17_lensing}
B.D.~Sherwin, A.~van Engelen, N.~Sehgal, M.~Madhavacheril, G.E.~Addison, S.~Aiola et~al., \emph{{Two-season Atacama Cosmology Telescope polarimeter lensing power spectrum}}, \href{https://doi.org/10.1103/PhysRevD.95.123529}{\emph{Phys. Rev. D} {\bfseries 95} (2017) 123529}.

\bibitem{Polarbear14}
{\scshape POLARBEAR} collaboration, \emph{{Measurement of the Cosmic Microwave Background Polarization Lensing Power Spectrum with the POLARBEAR Experiment}}, \href{https://doi.org/10.1103/PhysRevLett.113.021301}{\emph{Phys. Rev. Lett.} {\bfseries 113} (2014) 021301}.

\bibitem{BICEP18}
{The BICEP/Keck Collaboration}, P.A.R.~{Ade}, Z.~{Ahmed}, R.W.~{Aikin}, K.D.~{Alexander}, D.~{Barkats} et~al., \emph{{Measurements of Degree-Scale B-mode Polarization with the BICEP/Keck Experiments at South Pole}}, \href{https://doi.org/10.48550/arXiv.1807.02199}{\emph{arXiv e-prints} (2018) arXiv:1807.02199} [\href{https://arxiv.org/abs/1807.02199}{{\ttfamily 1807.02199}}].

\bibitem{Skalidis18}
{Skalidis, R.}, {Panopoulou, G. V.}, {Tassis, K.}, {Pavlidou, V.}, {Blinov, D.}, {Komis, I.} et~al., \emph{{Local measurements of the mean interstellar polarization at high Galactic latitudes}}, \href{https://doi.org/10.1051/0004-6361/201832827}{\emph{\aap} {\bfseries 616} (2018) A52}.

\bibitem{PlanckX16}
{Planck Collaboration}, R.~{Adam}, P.A.R.~{Ade}, N.~{Aghanim}, M.I.R.~{Alves}, M.~{Arnaud} et~al., \emph{{Planck 2015 results. X. Diffuse component separation: Foreground maps}}, \href{https://doi.org/10.1051/0004-6361/201525967}{\emph{\aap} {\bfseries 594} (2016) A10} [\href{https://arxiv.org/abs/1502.01588}{{\ttfamily 1502.01588}}].

\bibitem{PlanckXXX16_PIP_dust}
{Planck Collaboration}, R.~{Adam}, P.A.R.~{Ade}, N.~{Aghanim}, M.~{Arnaud}, J.~{Aumont} et~al., \emph{{Planck intermediate results. XXX. The angular power spectrum of polarized dust emission at intermediate and high Galactic latitudes}}, \href{https://doi.org/10.1051/0004-6361/201425034}{\emph{\aap} {\bfseries 586} (2016) A133} [\href{https://arxiv.org/abs/1409.5738}{{\ttfamily 1409.5738}}].

\bibitem{PlanckIV20_compsep}
{Planck Collaboration}, Y.~{Akrami}, M.~{Ashdown}, J.~{Aumont}, C.~{Baccigalupi}, M.~{Ballardini} et~al., \emph{{Planck 2018 results. IV. Diffuse component separation}}, \href{https://doi.org/10.1051/0004-6361/201833881}{\emph{\aap} {\bfseries 641} (2020) A4} [\href{https://arxiv.org/abs/1807.06208}{{\ttfamily 1807.06208}}].

\bibitem{PlanckXI20_dust}
{Planck Collaboration}, Y.~{Akrami}, M.~{Ashdown}, J.~{Aumont}, C.~{Baccigalupi}, M.~{Ballardini} et~al., \emph{{Planck 2018 results. XI. Polarized dust foregrounds}}, \href{https://doi.org/10.1051/0004-6361/201832618}{\emph{\aap} {\bfseries 641} (2020) A11} [\href{https://arxiv.org/abs/1801.04945}{{\ttfamily 1801.04945}}].

\bibitem{Krach16_fgs}
N.~{Krachmalnicoff}, C.~{Baccigalupi}, J.~{Aumont}, M.~{Bersanelli} and A.~{Mennella}, \emph{{Characterization of foreground emission on degree angular scales for CMB B-mode observations. Thermal dust and synchrotron signal from Planck and WMAP data}}, \href{https://doi.org/10.1051/0004-6361/201527678}{\emph{\aap} {\bfseries 588} (2016) A65} [\href{https://arxiv.org/abs/1511.00532}{{\ttfamily 1511.00532}}].

\bibitem{Delabrouille07}
J.~{Delabrouille} and J.F.~{Cardoso}, \emph{{Diffuse source separation in CMB observations}}, \href{https://doi.org/10.48550/arXiv.astro-ph/0702198}{\emph{arXiv e-prints} (2007) astro} [\href{https://arxiv.org/abs/astro-ph/0702198}{{\ttfamily astro-ph/0702198}}].

\bibitem{Leach08}
S.M.~{Leach}, J.F.~{Cardoso}, C.~{Baccigalupi}, R.B.~{Barreiro}, M.~{Betoule}, J.~{Bobin} et~al., \emph{{Component separation methods for the PLANCK mission}}, \href{https://doi.org/10.1051/0004-6361:200810116}{\emph{\aap} {\bfseries 491} (2008) 597} [\href{https://arxiv.org/abs/0805.0269}{{\ttfamily 0805.0269}}].

\bibitem{Abitbol21}
M.H.~{Abitbol}, D.~{Alonso}, S.M.~{Simon}, J.~{Lashner}, K.T.~{Crowley}, A.M.~{Ali} et~al., \emph{{The Simons Observatory: gain, bandpass and polarization-angle calibration requirements for B-mode searches}}, \href{https://doi.org/10.1088/1475-7516/2021/05/032}{\emph{\jcap} {\bfseries 2021} (2021) 032} [\href{https://arxiv.org/abs/2011.02449}{{\ttfamily 2011.02449}}].

\bibitem{Eriksen08}
H.K.~Eriksen, J.B.~Jewell, C.~Dickinson, A.J.~Banday, K.M.~Górski and C.R.~Lawrence, \emph{{Joint Bayesian Component Separation and CMB Power Spectrum Estimation}}, \href{https://doi.org/10.1086/525277}{\emph{The Astrophysical Journal} {\bfseries 676} (2008) 10}.

\bibitem{Stompor08}
R.~Stompor, S.~Leach, F.~Stivoli and C.~Baccigalupi, \emph{{Maximum likelihood algorithm for parametric component separation in cosmic microwave background experiments}}, \href{https://doi.org/10.1111/j.1365-2966.2008.14023.x}{\emph{Monthly Notices of the Royal Astronomical Society} {\bfseries 392} (2008) 216}.

\bibitem{Vacher22_moments}
L.~{Vacher}, J.~{Aumont}, L.~{Montier}, S.~{Azzoni}, F.~{Boulanger} and M.~{Remazeilles}, \emph{{Moment expansion of polarized dust SED: A new path towards capturing the CMB B-modes with LiteBIRD}}, \href{https://doi.org/10.1051/0004-6361/202142664}{\emph{\aap} {\bfseries 660} (2022) A111} [\href{https://arxiv.org/abs/2111.07742}{{\ttfamily 2111.07742}}].

\bibitem{Delabrouille09_nilc}
J.~{Delabrouille}, J.F.~{Cardoso}, M.~{Le Jeune}, M.~{Betoule}, G.~{Fay} and F.~{Guilloux}, \emph{{A full sky, low foreground, high resolution CMB map from WMAP}}, \href{https://doi.org/10.1051/0004-6361:200810514}{\emph{\aap} {\bfseries 493} (2009) 835} [\href{https://arxiv.org/abs/0807.0773}{{\ttfamily 0807.0773}}].

\bibitem{Vio08}
R.~{Vio} and P.~{Andreani}, \emph{{``Internal Linear Combination'' method for the separation of CMB from Galactic foregrounds in the harmonic domain}}, \href{https://doi.org/10.48550/arXiv.0811.4277}{\emph{arXiv e-prints} (2008) arXiv:0811.4277} [\href{https://arxiv.org/abs/0811.4277}{{\ttfamily 0811.4277}}].

\bibitem{Carones23_mcnilc}
A.~{Carones}, M.~{Migliaccio}, G.~{Puglisi}, C.~{Baccigalupi}, D.~{Marinucci}, N.~{Vittorio} et~al., \emph{{Multiclustering needlet ILC for CMB B-mode component separation}}, \href{https://doi.org/10.1093/mnras/stad2423}{\emph{\mnras} {\bfseries 525} (2023) 3117} [\href{https://arxiv.org/abs/2212.04456}{{\ttfamily 2212.04456}}].

\bibitem{Tegmark03}
M.~Tegmark, A.~de~Oliveira-Costa and A.J.S.~Hamilton, \emph{{High resolution foreground cleaned CMB map from WMAP}}, \href{https://doi.org/10.1103/PhysRevD.68.123523}{\emph{Phys. Rev. D} {\bfseries 68} (2003) 123523}.

\bibitem{Alonso17_forecasts}
D.~Alonso, J.~Dunkley, B.~Thorne and S.~N\ae{}ss, \emph{{Simulated forecasts for primordial $B$-mode searches in ground-based experiments}}, \href{https://doi.org/10.1103/PhysRevD.95.043504}{\emph{Phys. Rev. D} {\bfseries 95} (2017) 043504}.

\bibitem{Carones23_nilc}
{Carones, A.}, {Migliaccio, M.}, {Marinucci, D.} and {Vittorio, N.}, \emph{{Analysis of Needlet Internal Linear Combination performance on B-mode data from sub-orbital experiments}}, \href{https://doi.org/10.1051/0004-6361/202244824}{\emph{A\&A} {\bfseries 677} (2023) A147}.

\bibitem{BICEP14}
{\scshape BICEP2} collaboration, \emph{{Detection of $B$-Mode Polarization at Degree Angular Scales by BICEP2}}, \href{https://doi.org/10.1103/PhysRevLett.112.241101}{\emph{Phys. Rev. Lett.} {\bfseries 112} (2014) 241101}.

\bibitem{Tomita86}
H.~{Tomita}, \emph{{Curvature Invariants of Random Interface Generated by Gaussian Fields}}, \href{https://doi.org/10.1143/PTP.76.952}{\emph{Progress of Theoretical Physics} {\bfseries 76} (1986) 952}.

\bibitem{Coles87}
P.~{Coles} and J.D.~{Barrow}, \emph{{Non-Gaussian statistics and the microwave background radiation}}, \href{https://doi.org/10.1093/mnras/228.2.407}{\emph{\mnras} {\bfseries 228} (1987) 407}.

\bibitem{Gott90}
J.R.~{Gott}, III, C.~{Park}, R.~{Juszkiewicz}, W.E.~{Bies}, D.P.~{Bennett}, F.R.~{Bouchet} et~al., \emph{{Topology of Microwave Background Fluctuations: Theory}}, \href{https://doi.org/10.1086/168511}{\emph{\apj} {\bfseries 352} (1990) 1}.

\bibitem{Schmalzing98}
J.~{Schmalzing} and K.M.~{Gorski}, \emph{{Minkowski functionals used in the morphological analysis of cosmic microwave background anisotropy maps}}, \href{https://doi.org/10.1046/j.1365-8711.1998.01467.x}{\emph{\mnras} {\bfseries 297} (1998) 355} [\href{https://arxiv.org/abs/astro-ph/9710185}{{\ttfamily astro-ph/9710185}}].

\bibitem{PlanckVII20_isotropy}
{Planck Collaboration}, Y.~{Akrami}, M.~{Ashdown}, J.~{Aumont}, C.~{Baccigalupi}, M.~{Ballardini} et~al., \emph{{Planck 2018 results. VII. Isotropy and statistics of the CMB}}, \href{https://doi.org/10.1051/0004-6361/201935201}{\emph{\aap} {\bfseries 641} (2020) A7} [\href{https://arxiv.org/abs/1906.02552}{{\ttfamily 1906.02552}}].

\bibitem{Rahman21}
F.~Rahman, P.~Chingangbam and T.~Ghosh, \emph{{The nature of non-Gaussianity and statistical isotropy of the 408 MHz Haslam synchrotron map}}, \href{https://doi.org/10.1088/1475-7516/2021/07/026}{\emph{Journal of Cosmology and Astroparticle Physics} {\bfseries 2021} (2021) 026}.

\bibitem{Zurcher21}
D.~{Z{\"u}rcher}, J.~{Fluri}, R.~{Sgier}, T.~{Kacprzak} and A.~{Refregier}, \emph{{Cosmological forecast for non-Gaussian statistics in large-scale weak lensing surveys}}, \href{https://doi.org/10.1088/1475-7516/2021/01/028}{\emph{\jcap} {\bfseries 2021} (2021) 028} [\href{https://arxiv.org/abs/2006.12506}{{\ttfamily 2006.12506}}].

\bibitem{Ducout13}
A.~{Ducout}, F.R.~{Bouchet}, S.~{Colombi}, D.~{Pogosyan} and S.~{Prunet}, \emph{{Non-Gaussianity and Minkowski functionals: forecasts for Planck}}, \href{https://doi.org/10.1093/mnras/sts483}{\emph{\mnras} {\bfseries 429} (2013) 2104} [\href{https://arxiv.org/abs/1209.1223}{{\ttfamily 1209.1223}}].

\bibitem{Ganesan15}
V.~{Ganesan}, P.~{Chingangbam}, K.P.~{Yogendran} and C.~{Park}, \emph{{Primordial non-Gaussian signatures in CMB polarization}}, \href{https://doi.org/10.1088/1475-7516/2015/02/028}{\emph{\jcap} {\bfseries 2015} (2015) 028} [\href{https://arxiv.org/abs/1411.5256}{{\ttfamily 1411.5256}}].

\bibitem{PlanckXVI16_isotropy}
{Planck Collaboration}, P.A.R.~{Ade}, N.~{Aghanim}, Y.~{Akrami}, P.K.~{Aluri}, M.~{Arnaud} et~al., \emph{{Planck 2015 results. XVI. Isotropy and statistics of the CMB}}, \href{https://doi.org/10.1051/0004-6361/201526681}{\emph{\aap} {\bfseries 594} (2016) A16} [\href{https://arxiv.org/abs/1506.07135}{{\ttfamily 1506.07135}}].

\bibitem{Santos16}
L.~{Santos}, K.~{Wang} and W.~{Zhao}, \emph{{Probing the statistical properties of CMB B-mode polarization through Minkowski functionals}}, \href{https://doi.org/10.1088/1475-7516/2016/07/029}{\emph{\jcap} {\bfseries 2016} (2016) 029} [\href{https://arxiv.org/abs/1510.07779}{{\ttfamily 1510.07779}}].

\bibitem{CarronDuque24_MFs}
J.~{Carr{\'o}n Duque}, A.~{Carones}, D.~{Marinucci}, M.~{Migliaccio} and N.~{Vittorio}, \emph{{Minkowski Functionals in SO(3) for the spin-2 CMB polarisation field}}, \href{https://doi.org/10.1088/1475-7516/2024/01/039}{\emph{\jcap} {\bfseries 2024} (2024) 039} [\href{https://arxiv.org/abs/2301.13191}{{\ttfamily 2301.13191}}].

\bibitem{Carones24_MFs}
A.~{Carones}, J.~{Carr{\'o}n Duque}, D.~{Marinucci}, M.~{Migliaccio} and N.~{Vittorio}, \emph{{Minkowski functionals of CMB polarization intensity with PYNKOWSKI: theory and application to Planck and future data}}, \href{https://doi.org/10.1093/mnras/stad3002}{\emph{\mnras} {\bfseries 527} (2024) 756} [\href{https://arxiv.org/abs/2211.07562}{{\ttfamily 2211.07562}}].

\bibitem{Adler09}
R.~Adler and J.~Taylor, \emph{{Random Fields and Geometry}}, Springer Monographs in Mathematics, Springer New York (2009).

\bibitem{healpix}
K.M.~{G{\'o}rski}, E.~{Hivon}, A.J.~{Banday}, B.D.~{Wandelt}, F.K.~{Hansen}, M.~{Reinecke} et~al., \emph{{HEALPix: A Framework for High-Resolution Discretization and Fast Analysis of Data Distributed on the Sphere}}, \href{https://doi.org/10.1086/427976}{\emph{\apj} {\bfseries 622} (2005) 759} [\href{https://arxiv.org/abs/astro-ph/0409513}{{\ttfamily astro-ph/0409513}}].

\bibitem{healpy}
A.~{Zonca}, L.~{Singer}, D.~{Lenz}, M.~{Reinecke}, C.~{Rosset}, E.~{Hivon} et~al., \emph{{healpy: equal area pixelization and spherical harmonics transforms for data on the sphere in Python}}, \href{https://doi.org/10.21105/joss.01298}{\emph{The Journal of Open Source Software} {\bfseries 4} (2019) 1298}.

\bibitem{PySM}
B.~{Thorne}, J.~{Dunkley}, D.~{Alonso} and S.~{N{\ae}ss}, \emph{{The Python Sky Model: software for simulating the Galactic microwave sky}}, \href{https://doi.org/10.1093/mnras/stx949}{\emph{\mnras} {\bfseries 469} (2017) 2821} [\href{https://arxiv.org/abs/1608.02841}{{\ttfamily 1608.02841}}].

\bibitem{Zonca21_PySM}
A.~Zonca, B.~Thorne, N.~Krachmalnicoff and J.~Borrill, \emph{{The Python Sky Model 3 software}}, \href{https://doi.org/10.21105/joss.03783}{\emph{Journal of Open Source Software} {\bfseries 6} (2021) 3783}.

\bibitem{Panexp25_PySM}
{The Pan-Experiment Galactic Science Group}, J.~{Borrill}, S.E.~{Clark}, J.~{Delabrouille}, A.V.~{Frolov}, S.~{Ghosh} et~al., \emph{{Full-sky Models of Galactic Microwave Emission and Polarization at Sub-arcminute Scales for the Python Sky Model}}, \href{https://doi.org/10.48550/arXiv.2502.20452}{\emph{arXiv e-prints} (2025) arXiv:2502.20452} [\href{https://arxiv.org/abs/2502.20452}{{\ttfamily 2502.20452}}].

\bibitem{Basak11_nilc_temp}
S.~Basak and J.~Delabrouille, \emph{{A needlet internal linear combination analysis of WMAP 7-year data: estimation of CMB temperature map and power spectrum}}, \href{https://doi.org/10.1111/j.1365-2966.2011.19770.x}{\emph{Monthly Notices of the Royal Astronomical Society} {\bfseries 419} (2011) 1163}.

\bibitem{Wolz24}
{Wolz, K.}, {Azzoni, S.}, {Hervías-Caimapo, C.}, {Errard, J.}, {Krachmalnicoff, N.}, {Alonso, D.} et~al., \emph{{The Simons Observatory: Pipeline comparison and validation for large-scale B-modes}}, \href{https://doi.org/10.1051/0004-6361/202346105}{\emph{\aap} {\bfseries 686} (2024) A16}.

\bibitem{Narcowich06}
F.J.~Narcowich, P.~Petrushev and J.D.~Ward, \emph{{Localized Tight Frames on Spheres}}, \href{https://doi.org/10.1137/040614359}{\emph{SIAM Journal on Mathematical Analysis} {\bfseries 38} (2006) 574} [\href{https://arxiv.org/abs/https://doi.org/10.1137/040614359}{{\ttfamily https://doi.org/10.1137/040614359}}].

\bibitem{Pietrobon06}
D.~{Pietrobon}, A.~{Balbi} and D.~{Marinucci}, \emph{{Integrated Sachs-Wolfe effect from the cross correlation of WMAP 3year and the NRAO VLA sky survey data: New results and constraints on dark energy}}, \href{https://doi.org/10.1103/PhysRevD.74.043524}{\emph{\prd} {\bfseries 74} (2006) 043524} [\href{https://arxiv.org/abs/astro-ph/0606475}{{\ttfamily astro-ph/0606475}}].

\bibitem{Geller10}
D.~{Geller} and D.~{Marinucci}, \emph{{Spin Wavelets on the Sphere}}, \href{https://doi.org/10.1007/s00041-010-9128-3}{\emph{Journal of Fourier Analysis and Applications} {\bfseries 16} (2010) 840} [\href{https://arxiv.org/abs/0811.2935}{{\ttfamily 0811.2935}}].

\bibitem{CarronDuque19_ps}
J.~{Carr{\'o}n Duque}, A.~{Buzzelli}, Y.~{Fantaye}, D.~{Marinucci}, A.~{Schwartzman} and N.~{Vittorio}, \emph{{Point source detection and false discovery rate control on CMB maps}}, \href{https://doi.org/10.1016/j.ascom.2019.100310}{\emph{Astronomy and Computing} {\bfseries 28} (2019) 100310} [\href{https://arxiv.org/abs/1902.06636}{{\ttfamily 1902.06636}}].

\bibitem{Hamimeche_Lewis09}
S.~{Hamimeche} and A.~{Lewis}, \emph{{Properties and use of CMB power spectrum likelihoods}}, \href{https://doi.org/10.1103/PhysRevD.79.083012}{\emph{\prd} {\bfseries 79} (2009) 083012} [\href{https://arxiv.org/abs/0902.0674}{{\ttfamily 0902.0674}}].

\bibitem{Gerbino20_likelihood}
M.~{Gerbino}, M.~{Lattanzi}, M.~{Migliaccio}, L.~{Pagano}, L.~{Salvati}, L.~{Colombo} et~al., \emph{{Likelihood methods for CMB experiments}}, \href{https://doi.org/10.3389/fphy.2020.00015}{\emph{Frontiers in Physics} {\bfseries 8} (2020) 15} [\href{https://arxiv.org/abs/1909.09375}{{\ttfamily 1909.09375}}].

\bibitem{Lewis01_masked}
A.~Lewis, A.~Challinor and N.~Turok, \emph{{Analysis of CMB polarization on an incomplete sky}}, \href{https://doi.org/10.1103/PhysRevD.65.023505}{\emph{Phys. Rev. D} {\bfseries 65} (2001) 023505}.

\bibitem{Hartlap07}
{Hartlap, J.}, {Simon, P.} and {Schneider, P.}, \emph{{Why your model parameter confidences might be too optimistic. Unbiased estimation of the inverse covariance matrix}}, \href{https://doi.org/10.1051/0004-6361:20066170}{\emph{\aap} {\bfseries 464} (2007) 399}.

\bibitem{Allys19}
{Allys, E.}, {Levrier, F.}, {Zhang, S.}, {Colling, C.}, {Regaldo-Saint Blancard, B.}, {Boulanger, F.} et~al., \emph{{The RWST, a comprehensive statistical description of the non-Gaussian structures in the ISM}}, \href{https://doi.org/10.1051/0004-6361/201834975}{\emph{\aap} {\bfseries 629} (2019) A115}.

\bibitem{Blancard20}
{Regaldo-Saint Blancard, B.}, {Levrier, F.}, {Allys, E.}, {Bellomi, E.} and {Boulanger, F.}, \emph{{Statistical description of dust polarized emission from the diffuse interstellar medium - A RWST approach}}, \href{https://doi.org/10.1051/0004-6361/202038044}{\emph{\aap} {\bfseries 642} (2020) A217}.

\bibitem{Delouis22}
{Delouis, J.M.}, {Allys, E.}, {Gauvrit, E.} and {Boulanger, F.}, \emph{{Non-Gaussian modelling and statistical denoising of Planck dust polarisation full-sky maps using scattering transforms}}, \href{https://doi.org/10.1051/0004-6361/202244566}{\emph{\aap} {\bfseries 668} (2022) A122}.

\bibitem{Mousset2024}
L.~{Mousset}, E.~{Allys}, M.A.~{Price}, J.~{Aumont}, J.M.~{Delouis}, L.~{Montier} et~al., \emph{{Generative models of astrophysical fields with scattering transforms on the sphere}}, \href{https://doi.org/10.1051/0004-6361/202451396}{\emph{\aap} {\bfseries 691} (2024) A269} [\href{https://arxiv.org/abs/2407.07007}{{\ttfamily 2407.07007}}].

\bibitem{Campeti2025}
P.~{Campeti}, J.M.~{Delouis}, L.~{Pagano}, E.~{Allys}, M.~{Lattanzi} and M.~{Gerbino}, \emph{{From few to many maps: A fast map-level emulator for extreme augmentation of cosmic microwave background systematics datasets}}, \href{https://doi.org/10.1051/0004-6361/202554540}{\emph{\aap} {\bfseries 700} (2025) A136} [\href{https://arxiv.org/abs/2503.11643}{{\ttfamily 2503.11643}}].

\bibitem{PlanckIX18_ng}
{Planck Collaboration}, {Akrami, Y.}, {Arroja, F.}, {Ashdown, M.}, {Aumont, J.}, {Baccigalupi, C.} et~al., \emph{{Planck 2018 results - IX. Constraints on primordial non-Gaussianity}}, \href{https://doi.org/10.1051/0004-6361/201935891}{\emph{\aap} {\bfseries 641} (2020) A9}.

\end{thebibliography}\endgroup

\end{document}